\documentclass[a4paper,11pt]{article}
\pdfoutput=1 

\usepackage{jcappub} 

\usepackage[T1]{fontenc} 
\usepackage{graphicx} 
\usepackage{comment}

\newcommand{\mat}[1]{\boldsymbol{\mathsf{#1}}}

\title{The DESI DR1 Peculiar Velocity Survey: growth rate measurements from the maximum likelihood fields method}

\author[1, 2]{Y. Lai,}
\author[1]{C. Howlett,}
\author[3]{J.~Aguilar,}
\author[4]{S.~Ahlen,}
\author[5]{A.~J.~Amsellem,}
\author[6]{J. Bautista,}
\author[7]{S. BenZvi,}
\author[8, 9]{D.~Bianchi,}
\author[10]{C. Blake,}
\author[11]{D.~Brooks,}
\author[12]{A. Carr,}
\author[3]{T.~Claybaugh,}
\author[1]{T. M. Davis,}
\author[13]{A.~de la Macorra,}
\author[11]{P.~Doel,}
\author[7]{K. Douglass,}
\author[3, 14]{S.~Ferraro,}
\author[15]{A.~Font-Ribera,}
\author[16, 17]{J.~E.~Forero-Romero,}
\author[18, 19, 20]{E.~Gaztañaga,}
\author[21]{G.~Gutierrez,}
\author[3]{J.~Guy,}
\author[22, 23]{H.~K.~Herrera-Alcantar,}
\author[24, 25]{D.~Huterer,}
\author[26]{M.~Ishak,}
\author[27]{R.~Joyce,}
\author[3]{A. Kim,}
\author[28]{D.~Kirkby,}
\author[3]{T.~Kisner,}
\author[3]{A.~Kremin,}
\author[11]{O.~Lahav,}
\author[29]{C.~Lamman,}
\author[3]{M.~Landriau,}
\author[30]{L.~Le~Guillou,}
\author[31, 32]{A.~Leauthaud,}
\author[3]{M.~E.~Levi,}
\author[33, 15]{M.~Manera,}
\author[34, 35, 29]{P.~Martini,}
\author[27]{A.~Meisner,}
\author[36, 15]{R.~Miquel,}
\author[37]{J.~Moustakas,}
\author[13]{A.~Muñoz-Gutiérrez,}
\author[19]{S.~Nadathur,}
\author[38, 39, 40]{W.~J.~Percival,}
\author[3, 41, 14]{C.~Poppett,}
\author[42]{F.~Prada,}
\author[43]{I.~P\'erez-R\`afols,}
\author[6]{F. Qin,}
\author[1]{C. Ross,}
\author[44]{G.~Rossi,}
\author[1]{K. Said,}
\author[45]{E.~Sanchez,}
\author[3]{D.~Schlegel,}
\author[24, 25]{M.~Schubnell,}
\author[46]{H.~Seo,}
\author[3]{J.~Silber,}
\author[27]{D.~Sprayberry,}
\author[25]{G.~Tarl\'{e},}
\author[10]{R. Turner,}
\author[27]{B.~A.~Weaver,}
\author[30]{P.~Zarrouk,}
\author[3]{R.~Zhou,}
\author[47]{H.~Zou}

\affiliation[1]{School of Mathematics and Physics, The University of Queensland, QLD 4072, Australia}
\affiliation[2]{Department of Astronomy/Steward Observatory, The University of Arizona, 933 North Cherry Avenue, Tucson, AZ 85721, USA}
\affiliation[3]{Lawrence Berkeley National Laboratory, 1 Cyclotron Road, Berkeley, CA 94720, USA}
\affiliation[4]{Department of Physics, Boston University, 590 Commonwealth Avenue, Boston, MA 02215 USA}
\affiliation[5]{Department of Physics, Carnegie Mellon University, 5000 Forbes Avenue, Pittsburgh, PA 15213, USA}
\affiliation[6]{Aix-Marseille University, CNRS/IN2P3, CPPM, Marseille 13288, France}
\affiliation[7]{Department of Physics \& Astronomy, University of Rochester, 500 Joseph C. Wilson Blvd., Rochester, NY  14627, USA}
\affiliation[8]{Dipartimento di Fisica ``Aldo Pontremoli'', Universit\`a degli Studi di Milano, Via Celoria 16, I-20133 Milano, Italy}
\affiliation[9]{INAF-Osservatorio Astronomico di Brera, Via Brera 28, 20122 Milano, Italy}
\affiliation[10]{Centre for Astrophysics and Supercomputing, Swinburne University of Technology, P.O. Box 218, Hawthorn, VIC 3122, Australia}
\affiliation[11]{Department of Physics \& Astronomy, University College London, Gower Street, London, WC1E 6BT, UK}
\affiliation[12]{Korea Astronomy and Space Science Institute, 776 Daedeok-daero, Yuseong-gu, Daejeon 34055, South Korea}
\affiliation[13]{Instituto de F\'{\i}sica, Universidad Nacional Aut\'{o}noma de M\'{e}xico,  Circuito de la Investigaci\'{o}n Cient\'{\i}fica, Ciudad Universitaria, Cd. de M\'{e}xico  C.~P.~04510,  M\'{e}xico}
\affiliation[14]{University of California, Berkeley, 110 Sproul Hall \#5800 Berkeley, CA 94720, USA}
\affiliation[15]{Institut de F\'{i}sica d'Altes Energies (IFAE), The Barcelona Institute of Science and Technology, Edifici Cn, Campus UAB, 08193, Bellaterra (Barcelona), Spain}
\affiliation[16]{Departamento de F\'isica, Universidad de los Andes, Cra. 1 No. 18A-10, Edificio Ip, CP 111711, Bogot\'a, Colombia}
\affiliation[17]{Observatorio Astron\'omico, Universidad de los Andes, Cra. 1 No. 18A-10, Edificio H, CP 111711 Bogot\'a, Colombia}
\affiliation[18]{Institut d'Estudis Espacials de Catalunya (IEEC), c/ Esteve Terradas 1, Edifici RDIT, Campus PMT-UPC, 08860 Castelldefels, Spain}
\affiliation[19]{Institute of Cosmology and Gravitation, University of Portsmouth, Dennis Sciama Building, Portsmouth, PO1 3FX, UK}
\affiliation[20]{Institute of Space Sciences, ICE-CSIC, Campus UAB, Carrer de Can Magrans s/n, 08913 Bellaterra, Barcelona, Spain}
\affiliation[21]{Fermi National Accelerator Laboratory, PO Box 500, Batavia, IL 60510, USA}
\affiliation[22]{Institut d'Astrophysique de Paris. 98 bis boulevard Arago. 75014 Paris, France}
\affiliation[23]{IRFU, CEA, Universit\'{e} Paris-Saclay, F-91191 Gif-sur-Yvette, France}
\affiliation[24]{Department of Physics, University of Michigan, 450 Church Street, Ann Arbor, MI 48109, USA}
\affiliation[25]{University of Michigan, 500 S. State Street, Ann Arbor, MI 48109, USA}
\affiliation[26]{Department of Physics, The University of Texas at Dallas, 800 W. Campbell Rd., Richardson, TX 75080, USA}
\affiliation[27]{NSF NOIRLab, 950 N. Cherry Ave., Tucson, AZ 85719, USA}
\affiliation[28]{Department of Physics and Astronomy, University of California, Irvine, 92697, USA}
\affiliation[29]{The Ohio State University, Columbus, 43210 OH, USA}
\affiliation[30]{Sorbonne Universit\'{e}, CNRS/IN2P3, Laboratoire de Physique Nucl\'{e}aire et de Hautes Energies (LPNHE), FR-75005 Paris, France}
\affiliation[31]{Department of Astronomy and Astrophysics, UCO/Lick Observatory, University of California, 1156 High Street, Santa Cruz, CA 95064, USA}
\affiliation[32]{Department of Astronomy and Astrophysics, University of California, Santa Cruz, 1156 High Street, Santa Cruz, CA 95065, USA}
\affiliation[33]{Departament de F\'{i}sica, Serra H\'{u}nter, Universitat Aut\`{o}noma de Barcelona, 08193 Bellaterra (Barcelona), Spain}
\affiliation[34]{Center for Cosmology and AstroParticle Physics, The Ohio State University, 191 West Woodruff Avenue, Columbus, OH 43210, USA}
\affiliation[35]{Department of Astronomy, The Ohio State University, 4055 McPherson Laboratory, 140 W 18th Avenue, Columbus, OH 43210, USA}
\affiliation[36]{Instituci\'{o} Catalana de Recerca i Estudis Avan\c{c}ats, Passeig de Llu\'{\i}s Companys, 23, 08010 Barcelona, Spain}
\affiliation[37]{Department of Physics and Astronomy, Siena University, 515 Loudon Road, Loudonville, NY 12211, USA}
\affiliation[38]{Department of Physics and Astronomy, University of Waterloo, 200 University Ave W, Waterloo, ON N2L 3G1, Canada}
\affiliation[39]{Perimeter Institute for Theoretical Physics, 31 Caroline St. North, Waterloo, ON N2L 2Y5, Canada}
\affiliation[40]{Waterloo Centre for Astrophysics, University of Waterloo, 200 University Ave W, Waterloo, ON N2L 3G1, Canada}
\affiliation[41]{Space Sciences Laboratory, University of California, Berkeley, 7 Gauss Way, Berkeley, CA  94720, USA}
\affiliation[42]{Instituto de Astrof\'{i}sica de Andaluc\'{i}a (CSIC), Glorieta de la Astronom\'{i}a, s/n, E-18008 Granada, Spain}
\affiliation[43]{Departament de F\'isica, EEBE, Universitat Polit\`ecnica de Catalunya, c/Eduard Maristany 10, 08930 Barcelona, Spain}
\affiliation[44]{Department of Physics and Astronomy, Sejong University, 209 Neungdong-ro, Gwangjin-gu, Seoul 05006, Republic of Korea}
\affiliation[45]{CIEMAT, Avenida Complutense 40, E-28040 Madrid, Spain}
\affiliation[46]{Department of Physics \& Astronomy, Ohio University, 139 University Terrace, Athens, OH 45701, USA}
\affiliation[47]{National Astronomical Observatories, Chinese Academy of Sciences, A20 Datun Road, Chaoyang District, Beijing, 100101, P.~R.~China}

\emailAdd{ylai2@arizona.edu}
\emailAdd{c.howlett@uq.edu.au}

\abstract{We present the constraint on the growth rate of structure from the combination of DESI DR1 BGS sample, Fundamental Plane, and Tully-Fisher peculiar velocity catalogues using the maximum likelihood fields method. The combined catalogue contains 415,523 galaxy redshifts and 76,616 peculiar velocity measurements. To handle the large amount of data in the DESI DR1 peculiar velocity catalogue, we significantly improve the computational efficiency by rewriting the algorithm with \textsc{JAX}. After removing outliers and Tully-Fisher galaxies that are affected by systematics, we find \(f\sigma_8 = 0.483_{-0.043}^{+0.080}(\mathrm{stat}) \pm 0.018(\mathrm{sys})\), consistent within \(1\sigma\) with the power spectrum and correlation function analysis using the same dataset. Combining all three measurements with appropriate correlations, the consensus measurement is \(f\sigma_8 (z_{\mathrm{eff}}=0.07) = 0.450\pm0.055\), consistent with Planck \(+\Lambda\)CDM cosmology (\(f\sigma_8 = 0.449 \pm 0.008\)). Combining with the high redshift growth rate of structure measurements from DESI \textit{ShapeFit}, the constraint on the growth index is \(\gamma = 0.58\pm0.11\), consistent with GR.}

\begin{document}
\maketitle
\flushbottom

\section{Introduction}
Our concordance cosmological model, the \(\Lambda\) Cold Dark Matter model (\(\Lambda\)CDM), explains the accelerating expansion of the universe by introducing the cosmological constant into Einstein's theory of General Relativity (GR). The \(\Lambda\)CDM model has been the most popular cosmological model over the past 25 years and is supported by various observations \citep{Planck_2020, Brout_2022, Ivanov_2020, Percival_2007, Abbott_2022, Sugiyama_2023, Heymans_2021}. However, several internal inconsistencies start to appear. The most famous example is the Hubble tension, which is the discrepancy between measurements of the Hubble parameter from late-universe probes, such as supernovae \citep{Riess_2022}, and those from early-universe probes, such as the Cosmic Microwave Background \citep{Planck_2020}. Furthermore, DES (Dark Energy Survey) \citep{DES_2005} and DESI (Dark Energy Spectroscopic Instrument) \citep{Snowmass2013.Levi, DESI2016b.Instr, DESI2016a, DESI_Collaboration_2022, Guy_2023, Schlafly_2024, Abdul_Karim_2025, DESI_2025b, DESI_2025a, Miller_2023, Poppett_2024} found that the equation of state of dark energy may not be constant in time, contrary to the prediction of \(\Lambda\)CDM \citep{DES_2024, Adame_2025a, Adame_2025b, DESI_2025a, DESI_2025b}. An alternative explanation for the observed accelerating expansion is a modification to our theory of gravity (e.g., \citep{Dvali_2000, De_Felice_2010}). 

The strength of gravity varies across different theories of gravity \citep{Linder_2007}. This affects the distribution of large-scale structures in the late universe and the motions of galaxies induced by them. The rate at which these structures grow is characterised by the linear growth rate parameter \(f(a) = \frac{d \ln{\delta_m}}{d \ln{a}}\), where the scale factor \(a\) describes the relative size of the universe at different epochs and \(\delta_m\) is the matter overdensity. Hence, different gravity theories predict different linear growth rates at the same redshift. A model with weaker gravity than GR, such as the normal branch of the Dvali–Gabadadze–Porrati (nDGP) model \citep{Dvali_2000, Varun_Sahni_2003}, has a lower linear growth rate. On the other hand, a stronger gravity, such as the f(R) gravity models \citep{Hu_2007}, has a higher linear growth rate. Additionally, some modified gravity theories introduce scale dependence in the growth rate \citep{Sanchez_2010, Mirzatuny_2019, Denissenya_2022}. However, the differences in linear growth rates among various theories of gravity could be small, depending on the values of the additional degrees of freedom introduced by the model. Therefore, high-precision measurements of the linear growth rate are required to differentiate these theories.

We can constrain the linear growth rate of structure either by directly measuring the peculiar velocities of galaxies or by quantifying the change in the galaxy distribution inferred from redshifts (which are contaminated by the galaxies' peculiar velocities). The second effect is called Redshift Space Distortions (RSD; \citep{Jackson_1972, Kaiser_1986}). 

The peculiar velocity of a galaxy is the difference between its total velocity, which is related to its redshift, and its recession velocity from Hubble's law \citep{Davis_2014}. To calculate the recession velocity, we need the redshift-independent distance measurements, which are obtained via galaxy scaling relations, such as the Tully-Fisher relation \citep{Tully_1977} for spiral galaxies, the Fundamental Plane \citep{Djorgovski_1987, Dressler_1987} for elliptical galaxies, or standard candles such as Type Ia supernovae \citep{Phillips_1993}. In this work, we use the DESI DR1 peculiar velocity catalogue, which contains \(\sim 80,000\) peculiar velocity measurements using the Tully-Fisher relation \citep{DESIY1_TF} and Fundamental Plane \citep{DESIY1_FP}. It is the largest peculiar velocity catalogue to date and is almost twice as large as the Cosmicflows-4 sample \citep{Tully_2023}. 

In linear theory, the peculiar velocity and galaxy density are only sensitive to the parameter combinations  \(b\sigma_8\) and \(f\sigma_8\), where \(\sigma_8\) is the root mean square of matter density fluctuation within spheres of radius \(8h^{-1} \mathrm{Mpc}\), which defines the overall normalisation of the density perturbations, and \(b\) is the linear galaxy bias.\footnote{This degeneracy can be broken with the three-point correlation function/bispectrum or by combining the result with weak lensing \citep{Gil_Mar_n_2015, Massey_2007}.} The peculiar velocity and RSD are highly complementary methods to measure the growth rate of structure since peculiar velocity is more sensitive to the large-scale matter overdensity. At the same time, the RSD is more sensitive on nonlinear scales \citep{Koda_2014}. More importantly, the peculiar velocity and galaxy overdensity are two different tracers of the same underlying matter density field. Previous literature has shown that combining two different tracers of the same matter density field, such as galaxy density and peculiar velocity, can eliminate the sample variance and reduce the statistical uncertainty \citep{McDonald_2008, Blake_2013, Koda_2014}. 

Several methods have been developed to combine both tracers to constrain the growth rate. For example, we can determine the growth rate of structure by measuring the two-point velocity correlation functions \citep{Nusser_2017, Dupuy_2019, Turner_2021, Qin_2022, Turner_2022,Qin_2023a, Wang_2018, Beutler_2012, Appleby_2023}, measuring the galaxy density and peculiar velocity fields and constraining the growth rate by maximizing the likelihood \citep{Johnson_2014, Huterer_2017, Howlett_2017, Adams_2017, Adams_2020, Lai_2022}, measuring the momentum power spectrum \citep{Park_2000, Park_2006, Howlett_2019, Qin_2019, Fei_2025, Chen_2025}, and using the density field to reconstruct the predicted peculiar velocity field \citep{Carrick_2015, Boruah_2020, Said_2020, lilow_2021,Qin_2023, Boubel_2024, Stiskalek_2025}. In this work, we focus on the maximum likelihood fields method to constrain the growth rate of structure. 

The maximum likelihood fields method directly fits the correlation of the galaxy overdensity and peculiar velocity fields. Ref.~\citep{Adams_2017} were the first to simultaneously model the auto- and cross-covariance matrices of galaxy overdensities and peculiar velocities and constrain \(f\sigma_8\) by maximising the Gaussian likelihood function. They found that the constraint on the linear growth rate of structure is improved by 20\% compared to analysis using a single tracer. Ref.~\citep{Adams_2020} improved on this previous method by including the effect of small-scale RSD on the density field and velocity field. However, they assume that the lines of sight to all galaxies are parallel during the derivation. This assumption ignores the wide-angle effect. Previous literature has shown that this can introduce an error of more than 10\% when the subtended angle between the two line-of-sights exceeds 30 degrees \citep{Castorina_2018, Maresuke_2021}. Although ref.~\citep{Castorina_2020} developed a wide-angle formalism for the galaxy overdensity and peculiar velocity two-point correlation functions, they did not consider the damping in the galaxy and velocity power spectrum due to the effect of RSD on nonlinear scales (finger-of-god effect) \citep{Koda_2014}. Ref.~\citep{Lai_2022} accounts for both the wide-angle effect and the modelling of RSD on nonlinear scales by using the Taylor series approximation of the nonlinear RSD model. Ref.~\citep{Ravoux_2025} replace the Taylor series approximation with the Hankel transform to improve the accuracy and computational efficiency of the model.

Three different analysis pipelines in DESI are used to constrain the growth rate of structure with the DESI DR1 peculiar velocity catalogue. One companion paper~\citep{DESIY1_pk} constrains the growth rate of structure with the momentum power spectrum, while the other companion paper~\citep{DESIY1_xi} does the same with the correlation function. We demonstrate that, despite differences in methodology, these three approaches can obtain consistent constraints on the growth rate of structure. This consistency also allows us to combine the constraints to produce a consensus constraint on the growth rate of structure with the DESI DR1 peculiar velocity catalogue.  

This paper is organised as follows. In Section \ref{sec:Data}, we introduce the DESI DR1 peculiar velocity catalogue and its respective mock catalogues. In Section \ref{sec:model}, we briefly introduce the maximum likelihood fields method and highlight several changes that improve the computational efficiency of the algorithm. We present the results of tests on our methodology using mocks in Section \ref{sec:mock_test}. Section \ref{sec:result} then presents the constraint on the growth rate of structure from the DESI DR1 peculiar velocity catalogue and discusses its implications and comparison to previous constraints. Lastly, we present our conclusion in Section \ref{sec:conclusion}. Throughout this work, we assume a flat \(\Lambda\)CDM model with fiducial cosmological parameters given by \(\Omega_m = 0.3121\), \(H_0 = 100h \mathrm{km} \mathrm{s}^{-1} \mathrm{Mpc}^{-1} \) and \(\sigma_8 = 0.8150\) at redshift zero, consistent with the \(\Lambda\)CDM baseline cosmology in the \textsc{AbacusSummit} simulation \citep{Maksimova_2021, Garrison_2021, Hadzhiyska_2021}.  

\section{Data and simulation}
\label{sec:Data}
\subsection{The DESI DR1 peculiar velocity catalogue}
Dark Energy Spectroscopic Instrument (DESI) is a multi-object fibre spectrograph installed on the Mayall 4-meter telescope at Kitt Peak National Observatory to conduct a large-scale redshift survey over 8 years \citep{Snowmass2013.Levi, DESI2016b.Instr, DESI2016a}. It aims to collect approximately 63 million redshifts over the sky area of \(17,000 \mathrm{deg}^2\) below the redshift of four. The DESI peculiar velocity survey is a secondary targeting program of DESI that was designed to measure the peculiar velocities of galaxies in the local universe ($z \lesssim 0.15$) \citep{Saulder_2023}. DESI peculiar velocity survey uses both the Tully-Fisher \citep{Tully_1977} and the Fundamental Plane \citep{Dressler_1987, Djorgovski_1987} to measure the redshift-independent distances to spiral and elliptical galaxies, respectively. We refer readers to ref.~\citep{DESIY1_FP} and ref.~\citep{DESIY1_TF} for the detailed discussion of the construction of the DESI DR1 Fundamental Plane and Tully-Fisher catalogue, respectively. Here, we will mainly focus on the data product from these two catalogues. 

\begin{figure}
        \centering
	\includegraphics[width=1.0\textwidth]{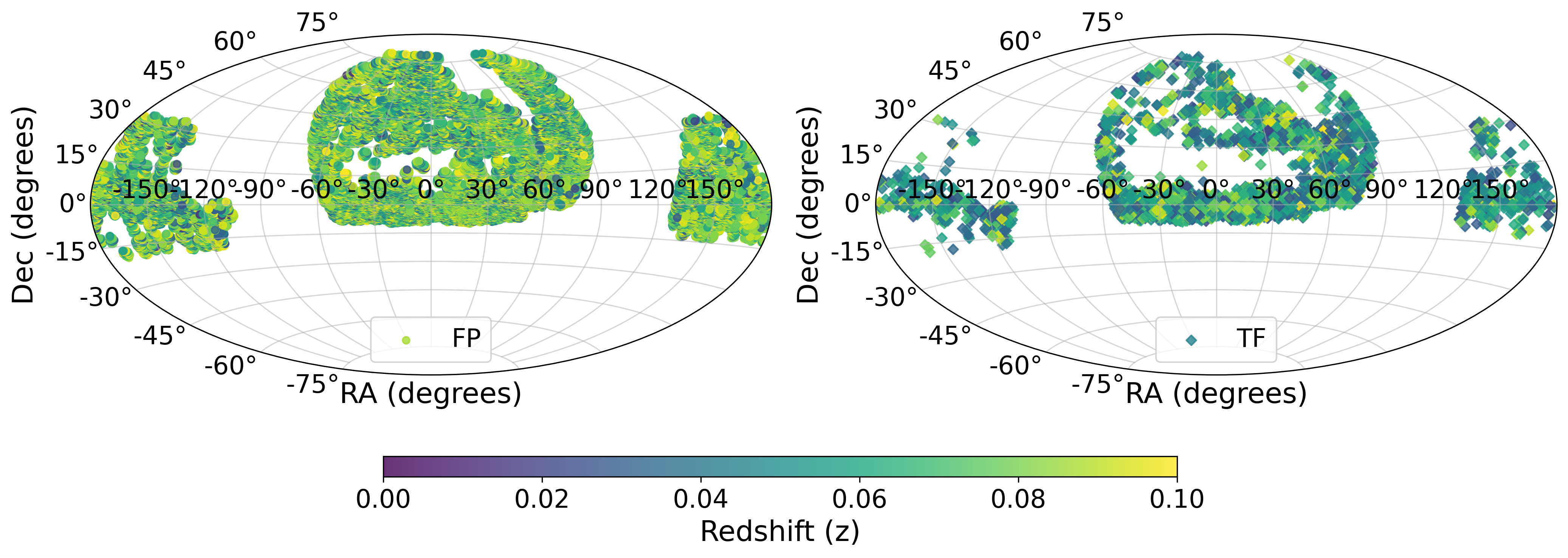}
    \caption{The distribution of Fundamental Plane (FP; circle, left) and Tully-Fisher (TF; diamond, right) galaxies in the DESI DR1 peculiar velocity catalogue. Their colours indicate the galaxies' redshifts. The DESI DR1 peculiar velocity catalogue mainly covers the northern hemisphere. }
    \label{fig:Distribution}
\end{figure}

The DESI DR1 Fundamental Plane catalogue \citep{DESIY1_FP} contains 73,822 elliptical galaxies between redshifts of \(0.01\) and \(0.1\) with a mean uncertainty in log-distance ratio of 0.12. Meanwhile, the Tully-Fisher catalogue \citep{DESIY1_TF} contains 6,806 spiral galaxies over the same redshift range, with a mean uncertainty in the log-distance ratio of 0.11. Fig.~\ref{fig:Distribution} illustrates the distribution of the Fundamental Plane galaxies (circle) and the Tully-Fisher galaxies (diamond) in the Aitoff projection. The colour of the marker indicates the galaxy's redshift. Most galaxies are in the Northern hemisphere, consistent with the DESI footprint. The patchy nature of the distribution is mainly due to the low completeness of the DR1 catalogue. This is expected to improve in future data releases. 

\begin{figure}
        \centering
	\includegraphics[width=0.7\textwidth]{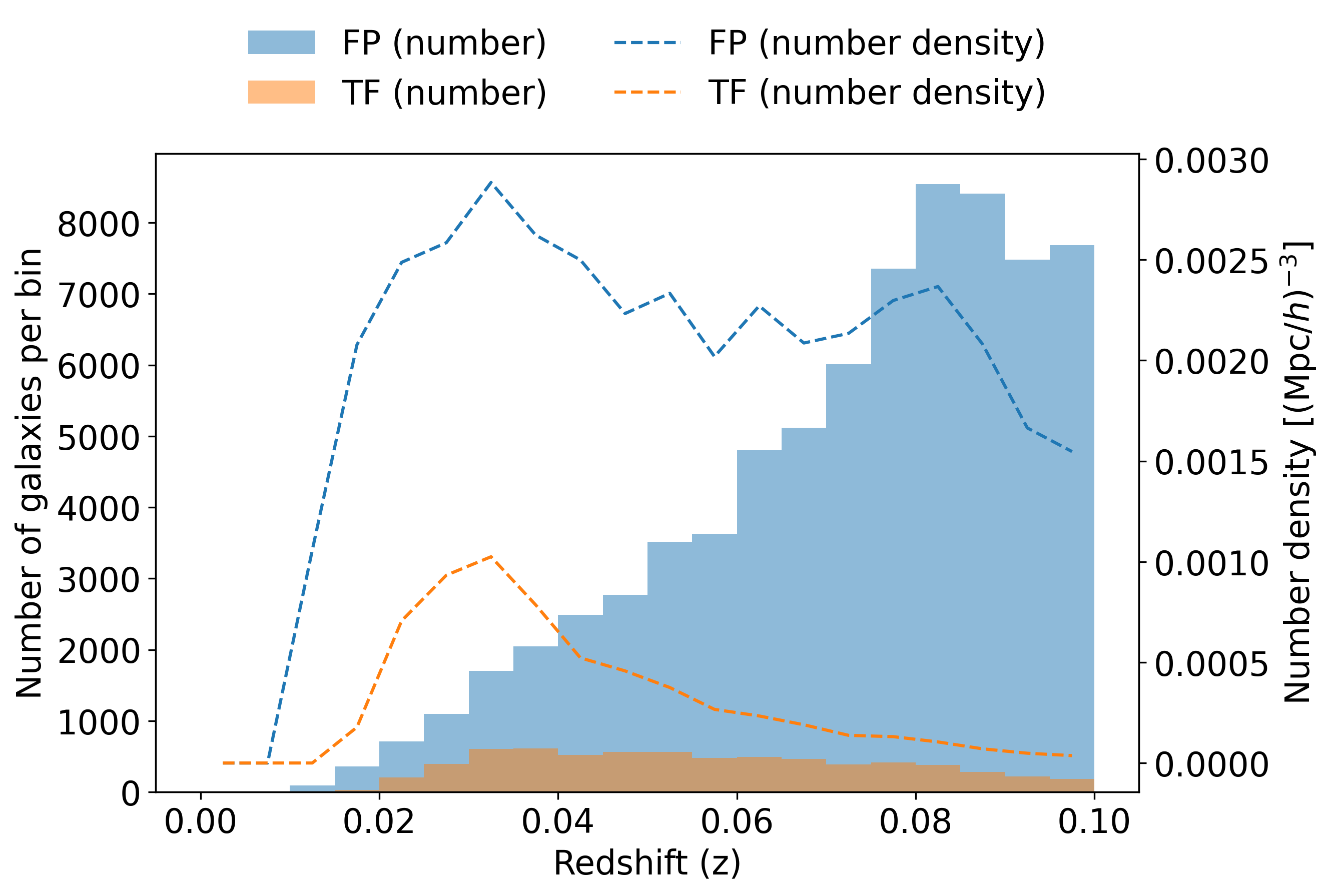}
    \caption{The redshift distribution of galaxies in the DESI DR1 Fundamental Plane and Tully-Fisher catalogue. The histogram shows the number of galaxies in each redshift bin, and the dashed line shows the number density of galaxies per unit volume in each redshift bin. The distribution of galaxies in the Fundamental Plane catalogue skews towards higher redshift, while the galaxies in the Tully-Fisher catalogue peak at low redshift. However, the number density of galaxies from both catalogues peaks at low redshift.}
    \label{fig:redshift_distribution}
\end{figure}

\begin{figure}
        \centering
	\includegraphics[width=0.495\textwidth]{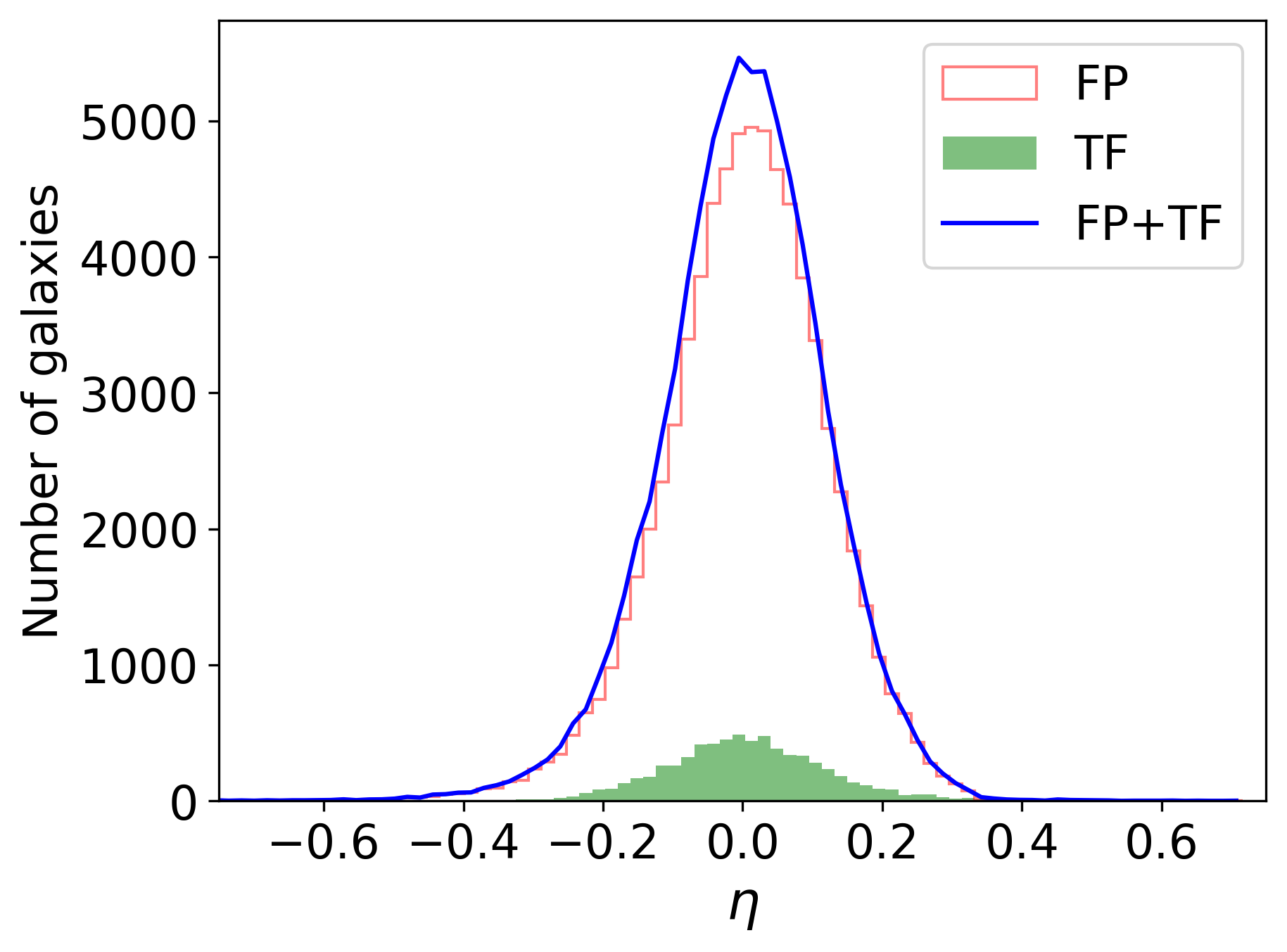}
        \includegraphics[width=0.495\textwidth]{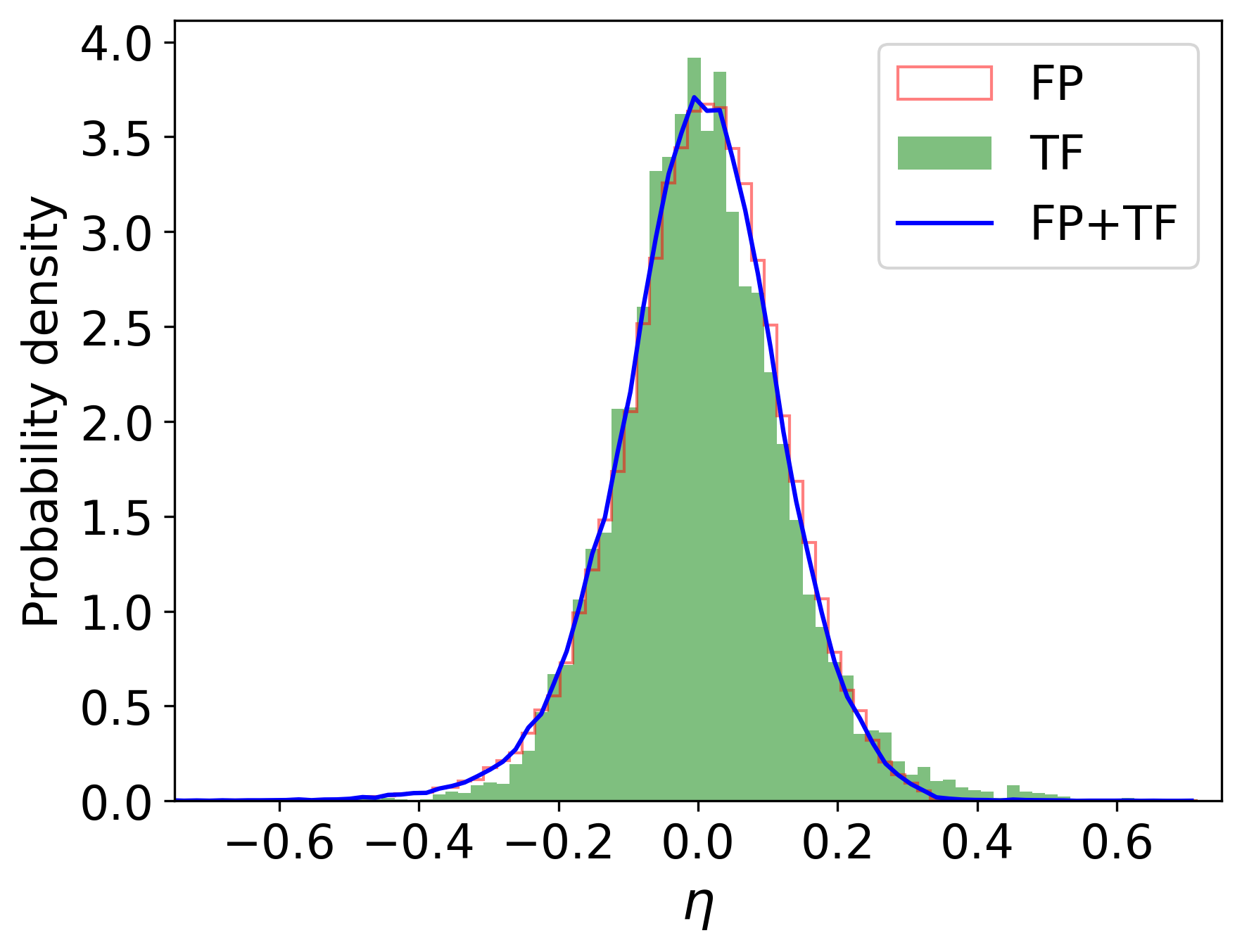}
    \caption{Distribution of the log-distance ratios \(\eta\) in the Fundamental Plane, Tully-Fisher, and the combined catalogues. The plot on the right is the normalised probability density. The Fundamental Plane catalogue skews towards a more negative log-distance ratio, while the Tully-Fisher catalogue skews towards a more positive log-distance ratio. }
    \label{fig:log_dist_ratio}
\end{figure}

Fig.~\ref{fig:redshift_distribution} illustrates the redshift distribution of the Fundamental Plane galaxies (blue) and Tully-Fisher galaxies (orange). The histogram shows the number of galaxies in each redshift bin, while the dashed line shows the number density of galaxies in each redshift bin. The distribution of Fundamental Plane galaxies skews towards higher redshift, while the number of Tully-Fisher galaxies peaks at low redshift. This is likely because elliptical galaxies are more massive, making them easier to observe at higher redshifts than spiral galaxies. Furthermore, to measure the rotation velocity of spiral galaxies, multiple fibres must be placed on the same galaxy. In contrast, the velocity dispersion of the elliptical galaxy can be measured with a single fibre. Additionally, the redshift bins at higher redshifts also have larger volumes, so they contain more galaxies. In contrast, the galaxy number density in both catalogues peaks at a redshift of 0.03. Fig.~\ref{fig:log_dist_ratio} illustrates the distribution of log-distance ratios in the Fundamental Plane, Tully-Fisher, and the combined catalogue. We normalised all three distributions in the figure on the right to better compare them. In general, the Fundamental Plane catalogue skews towards a more negative log-distance ratio with skewness of -0.53, while the Tully-Fisher catalogue skews towards a more positive log-distance ratio with skewness of 0.46. We do not expect the observed log-distance ratio to be perfectly Gaussian, since data systematics could alter the distribution's shape. Nonetheless, the skewness is small for both samples, so the log-distance-ratio distributions can be approximated by Gaussians. To further reduce the possibility that outliers may bias our constraints, we remove log-distance ratio measurements that are more than \(4\sigma\) away from the median. 

We found that the measurement of the growth rate from the Tully-Fisher catalogue is inconsistent with that from the Fundamental Plane catalogue when including galaxies beyond the redshift of 0.05 (see appendix \ref{sec:fit_choice} for detailed discussions). This result is consistent with the companion paper \citep{DESIY1_xi}, where we found excess correlation between the peculiar velocity measurements at high redshift from the Tully-Fisher catalogue compared to the Fundamental Plane catalogue, and compared to TF measurements at z<0.05. Since this is the first time the Tully-Fisher relation has been used to measure peculiar velocity beyond redshift 0.05 \citep{Kourkchi_2020}, previously unknown systematics could cause the inconsistency observed here. In section~\ref{sec:TF_mocks}, we find the Tully-Fisher mocks can return unbiased \(f\sigma_8\), so the bias is caused by systematics not captured by the mocks, such as the environmental dependence of the Tully-Fisher catalogue.  To establish a conservative measurement on the growth rate of structure, we decided to remove all peculiar velocity measurements in the Tully-Fisher catalogue with \(z>0.05\). This reduces the total number of galaxies to 76,756. However, we stress that this selection choice does not impact the constraint on the growth rate and conclusion of this work, as shown in appendix~\ref{sec:fit_choice}, since over 90\% of peculiar velocity measurements come from the Fundamental Plane galaxies.  

\subsection{The BGS density field sample}
In this work, we aim to analyse the combination of the peculiar velocity field with the galaxy density field at redshifts below 0.1 to improve constraints on cosmological parameters and reduce sample variance. For our galaxy sample, we use the DESI DR1 large-scale structure catalogues (LSS)\footnote{\url{https://data.desi.lbl.gov/doc/releases/dr1/}} \cite{Adame_2025d}. We use the lowest-redshift sample—the Bright Galaxy Sample (BGS) from DESI. The BGS sample is a magnitude-limited sample of galaxies with r-band magnitudes of 14 < $r$ < 19.5, extending to redshifts $z$ < 0.5 \citep{Hahn_2023}. We refer readers to ref.~\citep {Hahn_2023} for a more detailed discussion on the selection criteria and validation of the BGS sample. 

Here, we only use samples below the redshift of 0.1 to match the redshift range of the peculiar velocity sample. A further luminosity cut is applied to the r-band \(M_r-5\log_{10}h < 17.7\)
to produce a sample containing 415,523 galaxies \citep{DESIY1_xi}. We refer the reader to ref.~\citep{DESIY1_FP} for the construction of the data and random catalogues. 

\subsection{Mock catalogues}
The mock peculiar velocity catalogues are produced to closely replicate the data of the DESI DR1 peculiar velocity catalogue and allow us to test whether our methodology is unbiased before it is applied to the data. The 675 mocks are produced with the base \(\Lambda\)CDM cosmology from the \textsc{AbacusSummit} simulations \citep{Maksimova_2021, Garrison_2021, Hadzhiyska_2021}. There are 25 different realisations, and we subsample each realisation to 27 different mocks. The mocks are all generated at the snapshot redshift of 0.2 with the fiducial \(f\sigma_8 = 0.462\) calculated based on the fiducial cosmology\footnote{We used redshift of 0.2 since the value of \(f\sigma_8\) does not change greatly, and this snapshot redshift was used to build the DESI official BGS mock catalogues and has been extensively tested \citep{DESIY1_mock}.}. We refer the readers to ref.~\citep{DESIY1_mock} for more detailed information on the construction of the mocks. 

\section{Theoretical modelling}
\label{sec:model}
In the maximum likelihood fields method, we directly fit the galaxy overdensity and log-distance ratio, assuming a Gaussian likelihood function 
\begin{equation}
    P(\boldsymbol{S}|\boldsymbol{m}) = \frac{1}{\sqrt{(2\pi)^n|\boldsymbol{C}(\boldsymbol{m})|}} e^{\boldsymbol{S}^T \boldsymbol{C}(\boldsymbol{m})^{-1} \boldsymbol{S}}
    \label{eq:loglike_og}
\end{equation}
since on linear scales, both are expected to be Gaussian distributed. Here, \(\boldsymbol{C}\) is the covariance matrix, \(\boldsymbol{m}\) is an array of free parameters in the model, \(n\) is the number of free parameters and \(\boldsymbol{S} = (\boldsymbol{\delta_g, \boldsymbol{\eta}})\) is the data vector. To calculate the likelihood, the only unknown is the covariance matrix. Ref.~\citep{Lai_2022} derived the covariance matrix analytically, so we will not show the derivation here. In the next section, we show the final solution of the analytical covariance matrix. 

\subsection{The analytical covariance matrix}
The galaxy auto-covariance matrix is given by 
\begin{align}
        & \mat{C}_{gg} (\boldsymbol{s_1}, \boldsymbol{s_2}) = \sum_{p q} \frac{(-1)^{p+q}}{2^{p+q} p! q!} \sigma_g^{2(p+q)}
        \sum_{l} i^l   \biggl(b^2 \xi_{mm, l}^{p, q, 0}(s, 0)  H_{p, q}^{l}(\boldsymbol{s_1}, \boldsymbol{s_2})+ \notag \\
        & f^2 \xi_{\theta \theta, l}^{p, q, 0}(s,0) H_{p+1, q+1}^{l}(\boldsymbol{s_1}, \boldsymbol{s_2}) + bf \xi_{m\theta, l}^{p, q, 0}(s, 0)\biggl[H_{p+1, q}^{l}(\boldsymbol{s_1}, \boldsymbol{s_2}) + H_{p, q+1}^{l}(\boldsymbol{s_1}, \boldsymbol{s_2})\biggl]\biggl),
    \label{eq:gg_af}
\end{align}
where \(b\) is the linear galaxy bias, \(f\) is the logarithmic growth rate, and \(H_{p, q}^l\) is defined in equation~(\ref{eq:H}). We adopt the Gaussian model for damping in the galaxy power spectrum due to the nonlinear RSD \citep{Koda_2014}
\begin{equation}
    D_g = e^{-\frac{(k \mu \sigma_g)^2}{2}} = \sum_{i = 0}^{\infty} \frac{(-1)^i (k\sigma_g)^{2i}}{2^i i!} \mu^{2i},  
    \label{eq:D_g_Taylor}
\end{equation}
where \(\sigma_g\) is a free parameter that controls the strength of damping. The correlation function \(\xi_{ab, l}^{p, q, n}(r, \sigma_u)\) is given by 
\begin{equation}
    \xi_{ab, l}^{p, q, n}(r, \sigma_u) = \int_{k_{\mathrm{min}}}^{k_{\mathrm{max}}} \frac{k^2 dk}{2 \pi^2} P_{a b}(k) j_l(kr) k^{2(p+q)} D_u^n(k, \sigma_u),
    \label{eq:xi}
\end{equation}
where \(P_{ab}\) is the power spectrum of two fields, \(a\) and \(b\) and \(a, b \subset \{m, \theta\}\), where \(m\) denotes the matter density field and \(\theta\) denotes the velocity-divergence field. Different from ref.~\citep{Lai_2022}, we generate the power spectra with the Standard Perturbation Theory (SPT) \citep{Okumura_2014, Howlett_2019} instead of the Renormalised Perturbation Theory (RPT) \citep{Crocce_2006a, Crocce_2006b, Crocce_2008}. This is because the companion power spectrum \citep{DESIY1_pk} and the correlation function \citep{DESIY1_xi} papers are both developed based on the Standard Perturbation Theory. This helps us to better compare the three models. We will refer the readers to ref.~\citep{DESIY1_pk} for more details on the derivation of the power spectra in SPT. In Appendix~\ref{sec:SPT_vs_RPT}, we demonstrate that the constraint on \(f\sigma_8\) is unaffected by the change in the power spectrum model. The nuisance parameters account for the difference between SPT and RPT. Here, the damping in the velocity field due to the small scales RSD is given by \citep{Koda_2014}
\begin{equation}
    D_u(k, \sigma_u) = \frac{\sin(k\sigma_u)}{k\sigma_u},
    \label{eq:D_u}
\end{equation}
where \(\sigma_u\) is a free parameter that quantifies the strength of damping in the velocity power spectrum. We set \(k_{\mathrm{max}} = 0.20h \mathrm{Mpc}^{-1}\) where our quasi-linear model breaks down \citep{Koda_2014} and \(k_{\mathrm{min}} = 0.0002h \mathrm{Mpc}^{-1}\) since the DESI DR1 data may be affected by modes larger than the survey volume \citep{Adams_2017}.
The weighting function \(H_{p, q}^{l}\) is given by 
\begin{multline}
    H_{p, q}^{l}(\boldsymbol{s_1}, \boldsymbol{s_2}) = \sum_{l_1, l_2}\frac{4 \pi^2}{(2l_1+1)(2l_2 + 1)}\sum_{m, m_1, m_2} G_{m, m_1, m_2}^{l, l_1, l_2} 
    Y_{l, m}^*(\boldsymbol{s_1}-\boldsymbol{s_2}) Y_{l_1, m_1}^*(\boldsymbol{s_1}) Y_{l_2, m_2}^* (\boldsymbol{s_2}) a_{l_1}^{2p} a_{l_2}^{2q},
\label{eq:H}
\end{multline}
where \(Y\) denotes the spherical harmonics, \(*\) denotes the complex conjugate, \(a_l\) is the coefficient of the multipole decomposition of the line-of-sight angle \(\mu = \hat{\boldsymbol{r}} \cdot \hat{\boldsymbol{k}}\)
\begin{equation}
    a_{l}^{2n} = \frac{2l + 1}{2}\int_{-1}^{1} \mu^{2n}L_l(\mu)d\mu,
    \label{eq:multipole}
\end{equation}
and \(G\) denotes the Gaunt coefficient 
\begin{equation}
\begin{split}
    G^{L_1, L_2, L_3}_{M_1, M_2, M_3} = \sqrt{\frac{(2L_1+1)(2L_2+1)(2L_3+1)}{4 \pi}}
    \begin{pmatrix}
    L_1 & L_2 & L_3 \\
    0 & 0 & 0 
    \end{pmatrix}
    \begin{pmatrix}
    L_1 & L_2 & L_3 \\
    M_1 & M_2 & M_3
    \end{pmatrix},
\end{split}
\label{eq:Gaunt}
\end{equation}
where the matrices in equation~(\ref{eq:Gaunt}) represent the Wigner $3j$ symbols. We use \textsc{Mathematica} to evaluate the weighting function. Fig.~6 and 7 in ref.~\citep{Lai_2022} demonstrate that the weighting function controls the RSD effect of lines-of-sight to two different galaxies. It reaches a maximum when the two lines of sight are parallel to each other and a minimum when the two are perpendicular. 

We set the maximum of \(p, q\) to \(p_{\mathrm{max}} = q_{\mathrm{max}} = 3\) consistent with ref.~\citep{Lai_2022}. This is equivalent to using a third-order Taylor expansion approximation for equation~(\ref{eq:D_g_Taylor}). Since \(\sigma_g\) is on the order of unity \citep{Koda_2014}, \(k\sigma_g \lesssim 1\), the third-order Taylor expansion is accurate enough to calculate the analytical covariance matrix. Ref.~\citep{Ravoux_2025} developed the \textsc{flip} package\footnote{\url{https://github.com/corentinravoux/flip}} that can calculate the analytical covariance without the Taylor expansion in \(D_g\). Furthermore, it uses the parallelised Hankel transform to massively speed up the calculation of the analytical covariance matrix. However, varying \(\sigma_g\) will require precomputing the covariance matrix for different \(\sigma_g\) and interpolating them during MCMC, which is computationally expensive. Additionally, Figs.~6 to 8 in ref.~\citep{Ravoux_2025} illustrate that both pipelines provide a consistent covariance matrix for \(k_{\mathrm{max}} = 0.20 h\mathrm{Mpc}^{-1}\). Therefore, we decided to use the method described in ref.~\citep{Lai_2022} to constrain the growth rate with the DESI DR1 peculiar velocity catalogue. 

Similarly, the galaxy-velocity cross-covariance matrix is 
\begin{align}
        & \mat{C}_{gv} (s, \sigma_u) = (aHf) \sum_{p} \frac{(-1)^p}{2^p p!} \sigma_g^{2p} \sum_{l} i^{l+1} \biggl(\xi_{m\theta, l}^{p, -0.5, 1}(s, \sigma_u)H_{p, 0.5}^{l}(\boldsymbol{s_1}, \boldsymbol{s_2}) + \notag \\
        & f \xi_{\theta \theta, l}^{p, -0.5, 1} (s, \sigma_u)H_{p+1, 0.5}^{l}(\boldsymbol{s_1}, \boldsymbol{s_2})\biggl)
    \label{eq:gv_af}
\end{align}
and the velocity-galaxy cross-covariance matrix is the transpose of the galaxy-velocity cross-covariance matrix. Similar to the galaxy auto-covariance matrix, we set the maximum \(p\) to \(p_{\mathrm{max}} = 3\). Lastly, the velocity auto-covariance matrix is given by 
\begin{equation}
    \mat{C}_{vv} (s, \sigma_u) = (aHf)^2 \sum_{l} i^{l+2} \xi_{\theta \theta, l}^{-0.5, -0.5, 2}(s, \sigma_u) H_{0.5, 0.5}^{l}(\boldsymbol{s_1}, \boldsymbol{s_2}). 
    \label{eq:vv_af}
\end{equation}
The velocity auto-covariance matrix does not depend on \(\sigma_g\), so the highest order of \(l\) only depend on the highest combined exponent for \(\mu_1\) and \(\mu_2\), which is 2 \citep{Adams_2020, Lai_2022}. Therefore, the only two non-zero terms for the velocity auto-covariance matrix are \(l = 0\) and \(l = 2\). 

\subsection{Modification to the covariance matrix}
In the previous section, we presented the expressions for the analytical covariance matrices of the theoretical galaxy overdensity and peculiar velocity field. However, before fitting the data, we must ensure that the data and the model are being compared on equal footing and that the comparison is computationally feasible. In this section, we will discuss the modifications we applied to the analytical covariance matrix to address these two aspects. 
\subsubsection{Conversion between peculiar velocity and log-distance ratio}
Ref.~\citep{Johnson_2014, Springob_2014} found that the uncertainty of the peculiar velocity is log-normally distributed. To solve this problem, they introduce the log-distance ratio \(\eta\)
\begin{equation}
    \eta = \log_{10}\biggl(\frac{D(z_{\mathrm{obs}})}{D(z_H)}\biggl). 
    \label{eq:log-dist}
\end{equation}
Here, \(D\) is the comoving distance, \(z_{\mathrm{obs}}\) is the observed redshift, and \(z_H\) is the redshift converted from redshift-independent distance with the fiducial cosmology. Since we fit the log-distance ratio of the data catalogue, we need to convert the peculiar velocity auto- and cross-covariance matrices to the log-distance ratio auto- and cross-covariance matrices. We use \citep{Watkins_2015, Springob_2014} 
\begin{equation}
    \kappa(z_{\mathrm{obs}}) = \frac{1}{\ln{10}} \frac{1 + z_{\mathrm{obs}}}{D_z(z_{\mathrm{obs}}) H(z_{\mathrm{obs}})}
    \label{eq:convert}
\end{equation}
as the conversion factor. The auto- and cross-covariance matrices of the log-distance ratio are
\begin{equation}
    \mat{C}_{\eta \eta}(\boldsymbol{s_1}, \boldsymbol{s_2}, \sigma_u) = \kappa(z_{\boldsymbol{s_1}})\kappa(z_{\boldsymbol{s_2}}) \mat{C}_{vv}(\boldsymbol{s_1}, \boldsymbol{s_2}, \sigma_u),
    \label{eq:vv_eta}
\end{equation}

\begin{equation}
    \mat{C}_{\eta g}(\boldsymbol{s_1}, \boldsymbol{s_2}, \sigma_u) = \kappa(z_{\boldsymbol{s_1}}) \mat{C}_{vg}(\boldsymbol{s_1}, \boldsymbol{s_2}, \sigma_u),
    \label{eq:vg_eta}
\end{equation}
and
\begin{equation}
    \mat{C}_{g \eta}(\boldsymbol{s_1}, \boldsymbol{s_2}, \sigma_u) = \kappa(z_{\boldsymbol{s_2}}) \mat{C}_{gv}(\boldsymbol{s_1}, \boldsymbol{s_2}, \sigma_u).
    \label{eq:gv_eta}
\end{equation}
The galaxy auto-covariance matrix is not affected since it does not depend on the log-distance ratio. 

\subsubsection{Gridding correction}
The DESI BGS density sample contains \(\sim 420,000\) galaxies below the redshift of 0.1, and the DR1 peculiar velocity catalogue contains \(\sim 80,000\). This means a brute-force approach will require us to compute the inverse and determinant of a \(\sim 500,000\times 500,000 \) covariance matrix, which is computationally expensive and may be numerically unstable. To reduce the computational time, we follow ref.~\citep{Lai_2022} to reduce the dimensionality of the covariance matrix by gridding the data with a \(20h^{-1}\mathrm{Mpc}\) grid size. The centre of each grid is treated as $\boldsymbol{s}_{i}$ when calculating the covariance matrix. Lastly, gridding also helps smooth out the nonlinearity in the data, making it more appropriate to use the Gaussian likelihood function. 

The galaxy overdensity of each grid cell after gridding is given by 
\begin{equation}
    \delta_g^{\rm grid} = \frac{w - \alpha w_r}{\alpha w_r},
    \label{eq: delta_g}
\end{equation}
where the weight in the mock/data catalogue is 
\begin{equation}
    w = \sum_i^{N_g^\mathrm{grid}} w_i^{\mathrm{comp}} w_i^{\rm FKP}
    \label{eq: weight_grid}
\end{equation}
and \(w_r\) is the weight in the random catalogue of the same grid calculated with equation~(\ref{eq: weight_grid}). Here, \(w_i^{\mathrm{comp}}\) denotes the incompleteness weight of galaxy \(i\) and 
\begin{equation}
    w_i^{\rm FKP} = \frac{1}{1+\bar{n}_gP_0}
    \label{eq:FKP}
\end{equation}
is the FKP weight \citep{Feldman_1994}, 
\begin{equation}
    \alpha = \frac{\sum_j^{N_{g}}w_j}{\sum_k^{N_g^r}w_{k, r}} \approx \frac{N_g}{N_g^r}
    \label{eq:alpha}
\end{equation}
and \(N_g^\mathrm{grid}\) is the total number of galaxies inside the grid. Additionally, \(\bar{n}_g\) denotes the number density of galaxies, and we set the characteristic amplitude of the power spectrum \(P_0 = 1600 h^{-3}\mathrm{Mpc}^3\) following ref.~\citep{Fei_2025}. Lastly, \(N_g\) denotes the total number of galaxies in the mock/data catalogue and \(N_g^r\) denotes the total number of galaxies in the random catalogue. 

For the log-distance ratio, we choose not to apply the incompleteness weight because the missing galaxies may not have the same log-distance ratio as the observed galaxies. While we would expect two neighbouring measurements to share the same larger-scale velocity, in reality, the individual measurements would be dominated by noise, both physical (because two neighbouring galaxies in a cluster can have very different non-linear velocities) and observational. Simply duplicating an observation by applying the incompleteness weight may not be appropriate. Although we can calculate the FKP weight for the momentum power spectrum \citep{Howlett_2019}, it is not clear whether the optimal weight for the log-distance ratio will be the same.\footnote{We also tried to multiply the FKP weight for the momentum power spectrum by \(\kappa^2\) to convert it into the same unit as log-distance ratio. We applied this weight as an approximation of the FKP weight during the mock test and found it has a negligible impact on the constraint.} The FKP weight is designed to optimise the statistical uncertainty \citep{Feldman_1994}, so our uncertainty could potentially be slightly larger than the FKP weight is applied. Nonetheless, we don't expect it to significantly affect our result. Hence, we will assign no weight to the log-distance ratio when gridding.\footnote{We also implemented the inverse variance weight as an option in our pipeline. However, during mock testing, we also found this has little impact on the constraint.} Consequently, the mean log-distance ratio within each grid cell is given by \citep{Lai_2022} 
\begin{equation}
    \eta^{\rm grid} = \frac{1}{N_\eta^\mathrm{grid}}\sum_i^{N_\eta^\mathrm{grid}}\eta_i
    \label{eq:eta_mean}
\end{equation}
and the standard error of each grid is 
\begin{equation}
    \sigma_\eta = \frac{1}{N_\eta^\mathrm{grid}} \sqrt{\sum_i^{N_\eta^\mathrm{grid}}\sigma_{\eta, i}^2},
    \label{eq:sigma_eta}
\end{equation}
where \(N_\eta^\mathrm{grid}\) is the total number of log-distance ratio measurements within the grid.  

\begin{figure}
    \centering
	\includegraphics[width = 0.7\textwidth]{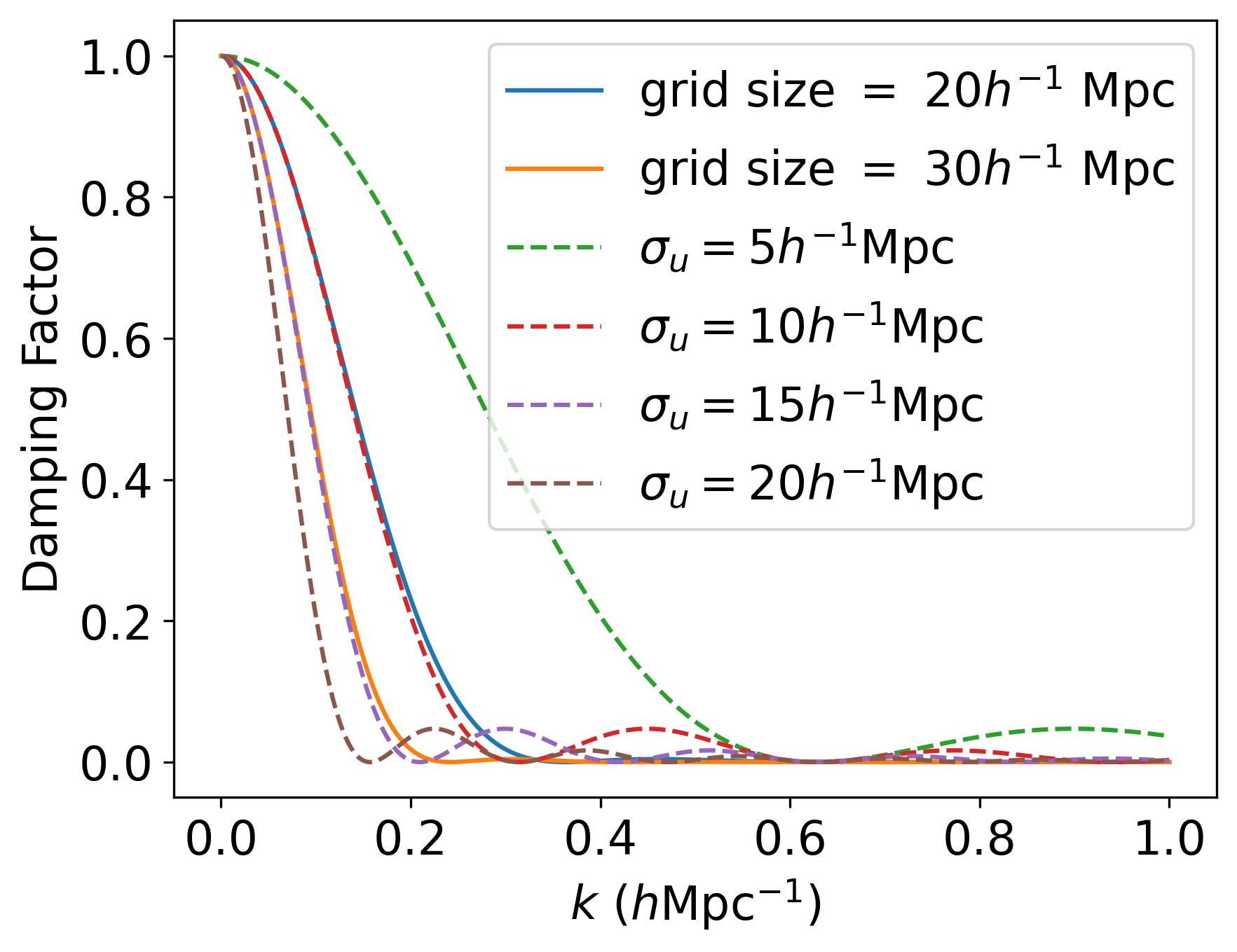}
    \caption{The damping on the power spectrum introduced by the gridding kernel and the RSD damping on the velocity field \(D_u\). The gridding kernel and the RSD damping are degenerate. The gridding kernel with a \(20h^{-1}\mathrm{Mpc}\) grid size is similar to applying the RSD damping with \(\sigma_u = 10h^{-1} \mathrm{Mpc}\). Similarly, the gridding kernel with a \(30h^{-1}\mathrm{Mpc}\) grid size is similar to applying the RSD damping with \(\sigma_u = 15h^{-1} \mathrm{Mpc}\).}
    \label{fig:damping_vs_grid_corr}
\end{figure}

Gridding removes most of the nonlinear information, so we have to modify the covariance matrix models accordingly. We can achieve this by replacing \citep{Johnson_2014}
\begin{equation}
   P_{XY}(k) \rightarrow P_{XY}^{\mathrm{grid}}(k) = P_{XY}(k) \Gamma(k)^2 
\end{equation}
when computing equation~(\ref{eq:xi}). $\Gamma(k)$ is the angle-averaged, Fourier transform of the gridding kernel with cell edge-length $L$\footnote{There is a typo in ref.~\citep{Howlett_2017, Lai_2022} for the expression of the gridding kernel, which misses the Jacobian \(\sin{\theta}\).} \citep{Johnson_2014}
\begin{equation}
     \Gamma(k) = \frac{1}{4 \pi} \int_0^{\pi} \int_0^{2 \pi} \sin{\theta} d\phi d\theta \frac{\sin{k_x} \sin{k_y} \sin{k_z}}{k_x k_y k_z}, 
     \label{eq:grid_window}
\end{equation}     
where the components of the wavevector are given by 
\begin{equation}
    k_x = \frac{k L}{2}\sin{\theta} \cos{\phi}; \quad k_y = \frac{k L}{2}\sin{\theta} \sin{\phi}; \quad k_z = \frac{k L}{2}\cos{\theta}. 
    \label{eq:k_x}
\end{equation} 
The gridding kernel is more important at small scales, where the wavenumber is comparable to or greater than the inverse of the grid size. Fig.~\ref{fig:damping_vs_grid_corr} illustrates the gridding kernel after numerically solving equation~(\ref{eq:grid_window}). The gridding kernel is degenerate with the RSD damping term that is applied to the velocity field. The gridding kernel with a \(20h^{-1}\mathrm{Mpc}\) grid size is similar to applying the RSD damping with \(\sigma_u = 10h^{-1} \mathrm{Mpc}\). Therefore, our constraint on \(\sigma_u\) could potentially be different from other approaches that do not need to apply the gridding correction. Appendix \ref{sec:fit_choice} discusses the effect of gridding on the \(f\sigma_8\) constraint.

Gridding also assumes that the galaxy densities and peculiar velocities are continuous. For the galaxy density, this is corrected by the shot noise (see Section~\ref{sec:shotnoise}). For the gridded version of the velocity auto-covariance matrix (\(C_{\eta \eta}^{\rm grid}\)), we follow ref.~\citep{Abate_2008} to update the diagonal elements with 
\begin{multline}
    \mat{C}_{\eta \eta}^{\mathrm{grid}}(\boldsymbol{s_1}, \boldsymbol{s_2}, \sigma_u) \rightarrow \mat{C}_{\eta \eta}^{\mathrm{grid}}(\boldsymbol{s_1}, \boldsymbol{s_2}, \sigma_u) + \frac{\mat{C}_{\eta \eta}(\boldsymbol{s_1}, \boldsymbol{s_2}, \sigma_u) - \mat{C}_{\eta \eta}^{\mathrm{grid}}(\boldsymbol{s_1}, \boldsymbol{s_2}, \sigma_u)}{N}\delta_{D}(\boldsymbol{s}_{1}-\boldsymbol{s}_{2})
    \label{eq:velocity_sn}
\end{multline}
to account for the difference introduced by averaging over different numbers of galaxies in different grid cells, where \(\mat{C}_{\eta \eta}\) is the covariance matrix without gridding. The correction is unnecessary for the off-diagonal component because it is negligible on small scales \citep{Abate_2008}. 

We further reduce the dimensionality of the covariance matrices by removing grid cells from the velocity auto- and cross-covariance matrices that contain no galaxy. Furthermore, we remove grid cells from the galaxy auto- and cross-covariance matrices for which no galaxy is present in the random catalogue. Both of these instances contribute no information to the likelihood function. To evaluate the impact of grid size on the \(f\sigma_8\) constraint, we decided to test both \(20 h^{-1}\mathrm{Mpc}\) grid size and \(30h^{-1}\mathrm{Mpc}\) grid size on the mocks.  

\subsubsection{Shot-noise correction}
\label{sec:shotnoise}
Both position and peculiar velocity of galaxies are treated as continuous fields in our model, although they are both discrete quantities. The error introduced is called the shot noise \(\sigma_{\delta_g}\) and can be modelled via Poisson statistics
\begin{equation}
    \sigma_{\delta_g}(x) = \frac{1}{\sqrt{N_{g}^r(x)}}. 
    \label{eq:shot-noise}
\end{equation}
We add the shot-noise contribution to the diagonal of the galaxy auto-covariance matrix
\begin{equation}
    \mat{C}_{gg}^{\mathrm{err}}(\boldsymbol{s_1}, \boldsymbol{s_2}, \sigma_g) = \mat{C}_{gg}^{\mathrm{grid}}(\boldsymbol{s_1}, \boldsymbol{s_2}, \sigma_g) + \sigma^{2}_{\delta_{g}}\delta_{D}(\boldsymbol{s}_{1}-\boldsymbol{s}_{2}).
    \label{eq:gg_sn}
\end{equation}
Similarly, the uncertainty of the log-distance ratio in each grid cell is also added to the corresponding diagonal of the covariance matrix
\begin{equation}
    \mat{C}_{\eta \eta}^{\mathrm{err}}(\boldsymbol{s_1}, \boldsymbol{s_2}, \sigma_u) = \mat{C}_{\eta \eta}^{\mathrm{grid}}(\boldsymbol{s_1}, \boldsymbol{s_2}, \sigma_u) + \sigma^{2}_{\eta}\delta_{D}(\boldsymbol{s}_{1}-\boldsymbol{s}_{2})
\end{equation}
and $\sigma_{\eta}^{\mathrm{grid}}$ contains two independent contributions to the uncertainty of the log-distance ratio $\sigma^{2}_{\eta} = (\sigma_{\eta}^{\mathrm{grid}})^{2} + \kappa(z_{\boldsymbol{s_1}})\kappa(z_{\boldsymbol{s_2}})\sigma_v^{2}$. The first term is the mean measurement uncertainty of the log-distance ratio for all galaxies in the grid cell, given by equation~(\ref{eq:sigma_eta}). The second term accounts for the velocity dispersion of galaxies on non-linear scales \(\sigma_v\), which is treated as a free parameter in our model.

\subsubsection{Integration bounds}
We follow ref.~\citep{Adams_2017, Lai_2022} to add an additional integral from \(k_{\mathrm{max}}\) to \(1.0 h\,\mathrm{Mpc^{-1}}\) for the galaxy auto-covariance matrix to account for the contribution to the galaxy auto-covariance matrix beyond \(0.2h\mathrm{Mpc}^{-1}\). This additional integral only acts as a nuisance parameter to increase the value of the galaxy-galaxy auto-covariance matrix, so it does not require complex modelling of the nonlinear power spectrum or the inclusion of redshift space distortions \citep{Adams_2020}. This additional covariance matrix is given by 
\begin{equation}
    \mat{C}_{gg}^{\mathrm{add}} =  \sum_{p q} \frac{(-1)^{p+q}}{2^{p+q} p! q!} \sigma_g^{2(p+q)}
        \sum_{l} i^l b_{\mathrm{add}}^2 \xi_{mm, l}^{p, q, 0}(r, 0) H_{p, q}^{l}, 
        \label{eq:b_add}
\end{equation}
where the integration bound for \(\xi\) is from \(k_{\mathrm{max}}\) to \(1.0h\,\mathrm{Mpc^{-1}}\) and \(b_{\mathrm{add}}\) is treated as a free parameter in our model. 
\subsubsection{Final covariance matrix and data vector}
After applying all the modifications to the covariance matrices outlined in the previous sections, the final covariance matrix is given by
\begin{equation}
\boldsymbol{\mathsf{C}} = 
\begin{pmatrix}
    \mat{C}_{gg} & \mat{C}_{gv} \\
    \mat{C}_{vg} & \mat{C}_{vv}
\end{pmatrix}
\rightarrow
\begin{pmatrix}
    \mat{C}_{gg}^{\mathrm{err}}+\mat{C}_{gg}^{\mathrm{add}} & \mat{C}_{g\eta}^{\mathrm{grid}} \\
    \mat{C}_{\eta g}^{\mathrm{grid}} & \mat{C}_{\eta \eta}^{\mathrm{err}}
\end{pmatrix}
.
\end{equation}
Similarly, the data vector is
\begin{equation}
\boldsymbol{S} = 
\begin{pmatrix}
    \boldsymbol{\delta}_{g} \\
    \boldsymbol{v}
\end{pmatrix}
\rightarrow
\begin{pmatrix}
    \boldsymbol{\delta}_{g}^{\mathrm{grid}} \\
    \boldsymbol{\eta}^{\mathrm{grid}}
\end{pmatrix}
.
\end{equation}

\subsection{Marginalisation over the zero-point correction}
The distributions of galaxy density and log-distance ratio are generally assumed to be Gaussian \citep{Adams_2017, Adams_2020}. However, one assumption of the fundamental plane and the Tully-Fisher relation is that the mean velocity within the sample is zero. The DESI peculiar velocity catalogue takes this into account by calibrating the peculiar velocity sample with Type Ia supernovae \citep{DESIY1_H0}. This procedure is called the zero-point correction. Ref.~\citep{Lai_2022} derived the likelihood function after marginalising over the zero-point correction uncertainty
\begin{equation}
    P(\boldsymbol{S}|m) = \frac{1}{\sqrt{(2\pi)^{n} |\mat{C(m)}|} N_{x}\sigma_{y}}e^{-\frac{1}{2}\left(\boldsymbol{S}^T \mat{C(m)}^{-1} \boldsymbol{S} -\frac{N_y^2}{N_x^2} \right)}
    \label{eq:MLF3}
\end{equation}
where \(N_x = \sqrt{\boldsymbol{x}^T \mat{C(m)}^{-1} \boldsymbol{x} + \frac{1}{\sigma_y^2}}\) and \(N_y = \boldsymbol{S}^T \mat{C(m)}^{-1} \boldsymbol{x}\). Here, \(\boldsymbol{x}\) has the same length as the data vector; it is zero for the galaxy density data and one for the log-distance ratio. This is because the zero-point correction does not affect the galaxy density. Lastly, the uncertainty of the zero-point correction for the DESI peculiar velocity catalogue is  \(\sigma_y = 0.005\) \citep{DESIY1_H0}.

\subsection{Improving computational efficiency}
Ref.~\citep{Lai_2022} showed that with the SDSS peculiar velocity catalogue, directly calculating the logarithmic likelihood could take around 60 seconds with a \(\sim 5000 \times 5000\) covariance matrix. Consequently, it may take weeks for the chains to converge using Markov Chain Monte Carlo (MCMC) algorithms. Ref.~\citep{Lai_2022} circumvents this issue by approximating the likelihood using a second-order Taylor expansion around 50 fiducial points evenly distributed across the prior of \(f\sigma_8\), while fixing other nuisance parameters to their best-fit values. These best fits are evaluated with the optimisation algorithm, which requires a much smaller number of iterations than the MCMC. Furthermore, the Taylor expansion of the likelihood depends on the score and Hessian matrix of the likelihood, whose dimensions are given by the number of free parameters in the model. This speeds up MCMC by more than a factor of 10 and could return the constraint on \(f\sigma_8\) within a couple of hours. The Taylor expansion will also slightly shift the mean of the posterior of \(f\sigma_8\) and changes its uncertainty as shown in Fig.~8 of ref.~\citep{Lai_2022}. However, such changes are much less than the uncertainty of \(f\sigma_8\) in ref.~\citep{Lai_2022}, so it was dismissed. In contrast, DESI peculiar velocity catalogues and the BGS sample are predicted to obtain a \(\sim 5\%\) \citep{Saulder_2023} constraint on \(f\sigma_8\), almost four times tighter than that from ref.~\citep{Lai_2022} (18.2\%). Therefore, using the Taylor expansion could potentially introduce significant systematics. We implemented several changes to improve the computational efficiency of exact logarithmic-likelihood calculation. 

We rewrote our code using \textsc{JAX} \citep{jax_2018} in \textsc{Python}, as \textsc{JAX} enables Just-In-Time (JIT) and XLA (Accelerated Linear Algebra) compilers to significantly speed up numerical calculations, particularly those involving linear algebra. We additionally sped up the likelihood calculation by applying the Cholesky decomposition. During our test with the mock catalogue, we found that the combination of \textsc{Jax} and Cholesky decomposition can reduce the likelihood computation time for a \(\sim 11000 \times 11000\) covariance matrix to around 1 second per evaluation. Lastly, we also replaced the MCMC sampling algorithm from \textsc{emcee} \citep{Foreman_Mackey_2013} to the Adaptive Sampling Metropolis-Hastings algorithm \citep{Zhu_2019} within \textsc{Numpyro} \citep{Phan_2019, Bingham_2019}. We consider the MCMC has converged if the Gelman-Rubin statistics of \(f\sigma_8\) reaches \(|R-1| \le 0.01\). It takes around 40,000 iterations for the chain to converge so that the analysis can be finished within a day.

\section{Testing on the DESI DR1 mocks}
\label{sec:mock_test}
In this section, we demonstrate the robustness of our theoretical model and fitting methodology using the DESI DR1 peculiar velocity mock catalogues, which are created based on a known cosmological model. We also use these fits to quantify fiducial values for parameters not varied in our data fitting, our systematic error budget, and the expected statistical error for our data. A more rigorous test requires us to use the approximate Neyman construction in ref.~\citep{Armstrong_2023} to assess the consistency in the confidence interval of our constraints. However, there are no peculiar velocity mocks available that are generated with a different cosmology, so we will leave this for future work.  

We have 675 different mocks, and MCMC is computationally expensive, so we decided to find the best-fit \(f\sigma_8\) of each mock through numerical optimisation with \textsc{JAXopt} \citep{Blondel_2022} and estimate the uncertainty with its Fisher matrix.\footnote{Estimated as the negative of the Hessian matrix at the maximum likelihood. The analytical expression of the Hessian matrix and the score function are provided in ref.~\citep{Lai_2022}.} Compared to the MCMC, which takes around 20 hours to finish analysing a combined dataset with both galaxy density and peculiar velocity, the optimisation will only take around 15 minutes. To ensure the optimisation returns robust results, we decided that the optimisation is successful if either the first derivative (score function) is less than \(0.01\) for all free parameters or the relative change in the best-fit parameters between two consecutive iterations is less than \(1\%\). If neither condition is met after 10 iterations, we deem the optimisation unsuccessful and exclude them from our final result. Except for the Tully-Fisher mocks (\(\sim 71\%\) success rate due to the low number of galaxies), which have much less constraining power, over 99\% of optimisations are successful. In this section, the ``mean" (red dot in Fig.~\ref{fig:mock_TF}, \ref{fig:mock_FP}, \ref{fig:mock_combined_pv}, and \ref{fig:mock_combined}) is estimated by taking the average of best-fit \(f\sigma_8\) from \textsc{JAXopt} with all the mocks, the ``mean uncertainty" (width of the red line is Fig.~\ref{fig:mock_TF}, \ref{fig:mock_FP}, \ref{fig:mock_combined_pv}, and \ref{fig:mock_combined}) is the mean of all mock uncertainties estimated from the Fisher matrix, the ``mock standard deviation" (black dashed line in Fig.~\ref{fig:mock_TF}, \ref{fig:mock_FP}, \ref{fig:mock_combined_pv}, and \ref{fig:mock_combined}) is approximated as the standard deviation of \(f\sigma_8\) from all mocks around the mock mean\footnote{The mock standard deviation is the combination of the statistical uncertainty and the sample variance since the best-fit \(f\sigma_8\) is also affected by the statistical limitation of the method.}, and the ``relative uncertainty" is the uncertainty estimated from the Fisher matrix divided by the best-fit \(f\sigma_8\) from \textsc{JAXopt}\footnote{The y-axis value of mock mean (red line) in Fig.~\ref{fig:mock_TF}, \ref{fig:mock_FP}, \ref{fig:mock_combined_pv}, and \ref{fig:mock_combined} is calculated by dividing the mean uncertainty of all mocks by their mean best-fit \(f\sigma_8\).}. 

\subsection{Free parameters}
There are six different free parameters in our model: \(f\sigma_8, \sigma_v, b\sigma_8, b_{\mathrm{add}}\sigma_8, \sigma_u\) and \(\sigma_g\).\footnote{Different from ref.~\citep{Adams_2020}, we do not treat \(k_{\mathrm{max}}\) as a free parameter. This is because Fig.~\ref{fig:damping_vs_grid_corr} demonstrates that the damping function strongly suppresses the power spectrum on small scales. Therefore, changing \(k_{\mathrm{max}}\) does not significantly affect the covariance matrix and the constraint on \(f\sigma_8\) as shown in ref.~\citep{Lai_2022}.} Of these, only \(\sigma_u\) cannot be varied by simply rescaling a pre-computed version of the appropriate part of the full covariance matrix model with the fiducial cosmological parameters \citep{Lai_2022}. Therefore, we need to compute the analytical covariance matrix for each MCMC iteration, which is computationally expensive. To reduce the computational time, we fix \(\sigma_u\) and justify the fiducial value based on our fits to the mocks. The change in \(f\sigma_8\) due to the change of \(\sigma_u\) will be included as the systematic uncertainty.

We assign flat priors for the remaining five free parameters based on previous works \citep{Johnson_2014, Howlett_2017_b, Adams_2017, Adams_2020, Lai_2022}. The prior for the normalized linear growth rate is \(0 \leq f\sigma_8 \leq 1\), the prior for the normalized galaxy bias is \(0 \leq b\sigma_8 \leq 3\), the prior for the nonlinear velocity dispersion is \(1 \mathrm{km s^{-1}}\) \(\leq \sigma_v \leq 2000 \mathrm{km s^{-1}}\), the prior for the additional galaxy bias is \(0 \leq b_{\mathrm{add}}\sigma_8 \leq 10\) and the prior for the finger-of-god damping term is \(0h^{-1} \mathrm{Mpc}\) \(\leq \sigma_g \leq\) \(10h^{-1} \mathrm{Mpc}\).

\subsection{Results from testing the Tully-Fisher mocks}
\label{sec:TF_mocks}
\begin{figure}
        \centering
	\includegraphics[width=1.0\textwidth]{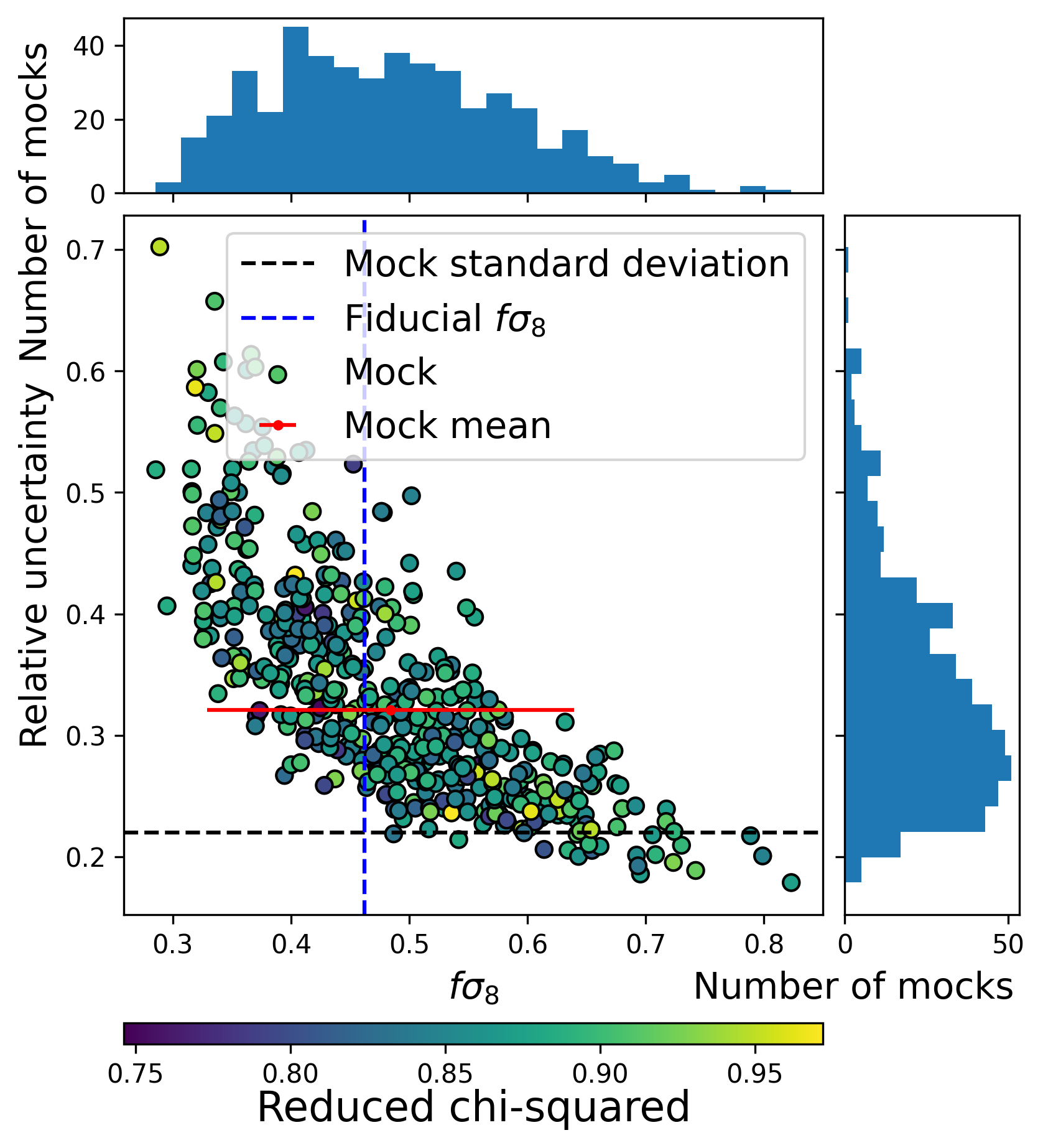}
    \caption{The constraints on \(f\sigma_8\) from 479 Tully-Fisher mocks with \(\sigma_u = 10h^{-1} \mathrm{Mpc}\) and grid size of \(20h^{-1} \mathrm{Mpc}\). The colour indicates the reduced chi-squared of each mock. The mock mean is consistent with the fiducial \(f\sigma_8\) indicated by the blue dashed line. }
    \label{fig:mock_TF}
\end{figure}

Fig.~\ref{fig:mock_TF} illustrates the constraints on \(f\sigma_8\) from 479 Tully-Fisher mocks with \(\sigma_u = 10h^{-1} \mathrm{Mpc}\) and grid size of \(20h^{-1} \mathrm{Mpc}\).  Fig.~\ref{fig:mock_TF} illustrates that the mock mean \(f\sigma_8 = 0.484\) is consistent with the fiducial value \(f\sigma_8=0.462\) indicated by the blue dashed line. Therefore, setting \(\sigma_u = 10h^{-1} \mathrm{Mpc}\) and grid size of \(20 h^{-1} \mathrm{Mpc}\) returns mean \(f\sigma_8\) closest to the fiducial value. The reduced chi-squared value for most constraints is around 0.9, indicating that the model is a good fit to the mock data. Therefore, the same grid size and \(\sigma_u\) should be chosen when only fitting the Tully-Fisher data. Additionally, the relative uncertainty decreases with increasing \(f\sigma_8\). Since the uncertainty on the log-distance is similar, a mock with a higher \(f\sigma_8\) will also have a higher signal-to-noise, resulting in a lower relative uncertainty. Lastly, the standard deviation (0.102) of the best-fit \(f\sigma_8\) is lower than the mean uncertainty (0.155), indicating we are not underestimating the uncertainty.\footnote{The difference between the standard deviation and the mean uncertainty here is likely because our likelihood function is a combination of a Gaussian and a non-Gaussian part, arising from marginalising over the zero-point correction. Since the Tully-Fisher catalogue contains fewer galaxies, the non-Gaussian parts dominate. The Fisher matrix only provides symmetry uncertainty, so it could cause the discrepancy here. A rigorous approach with MCMC is computationally expensive, and we do not provide constraints on cosmological parameters with only the Tully-Fisher catalogue.}      

\subsection{Results from testing the Fundamental Plane mocks}
\begin{figure}
        \centering
	\includegraphics[width=1.0\textwidth]{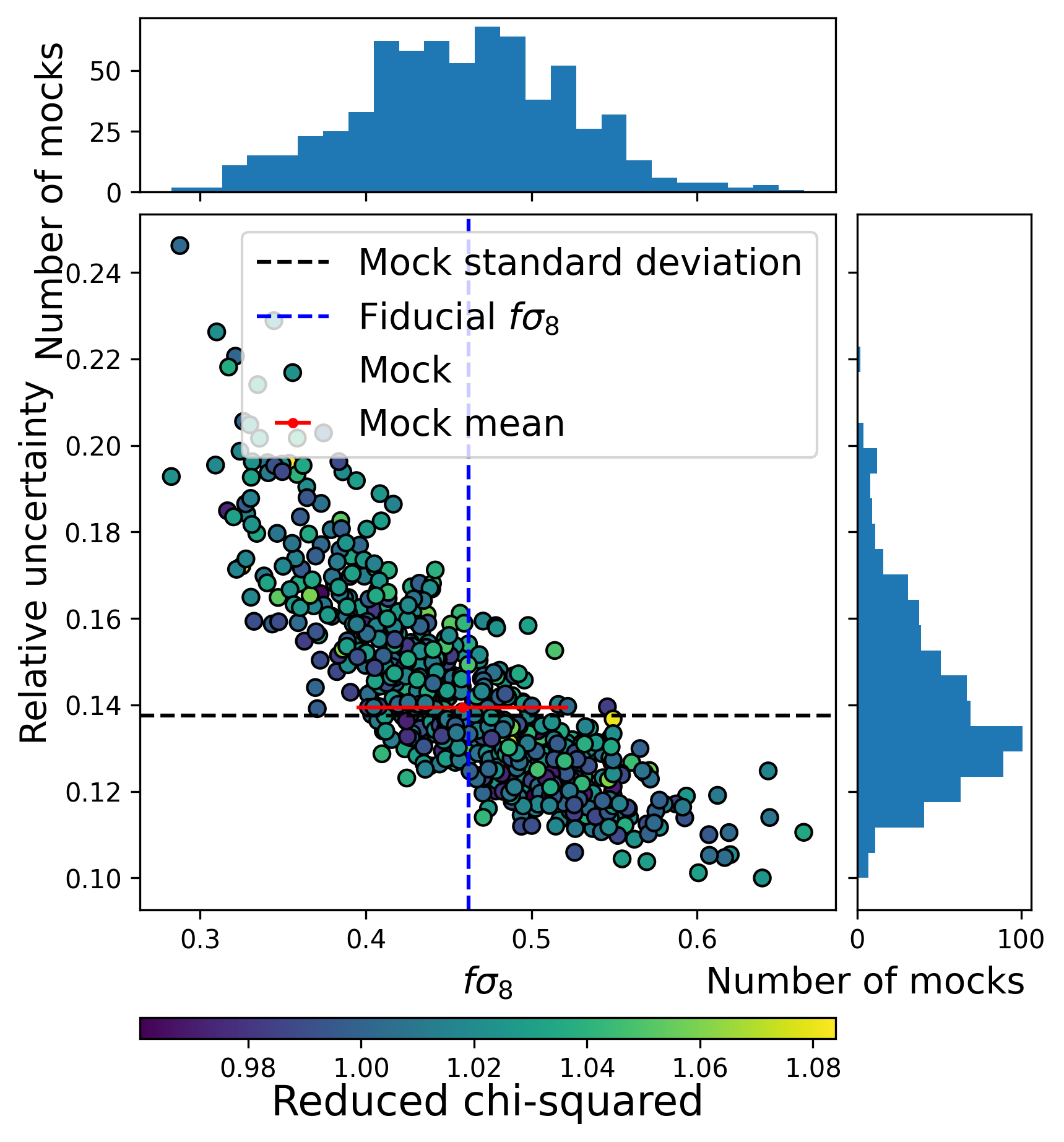}
    \caption{The constraints on \(f\sigma_8\) from 674 Fundamental Plane mocks with \(\sigma_u = 10h^{-1} \mathrm{Mpc}\) and grid size of \(20h^{-1} \mathrm{Mpc}\). The colour indicates the reduced chi-squared of each mock. The mock mean is consistent with the fiducial \(f\sigma_8\) indicated by the blue dashed line. }
    \label{fig:mock_FP}
\end{figure}
Fig.~\ref{fig:mock_FP} illustrates the constraints on \(f\sigma_8\) from 674 Fundamental Plane mocks with \(\sigma_u = 10h^{-1} \mathrm{Mpc}\) and grid size of \(20h^{-1} \mathrm{Mpc}\). The mock mean \(f\sigma_8 = 0.458\) is consistent with the fiducial value of \(f\sigma_8 = 0.462\) indicated by the blue dashed line, so our configuration returns an unbiased constraint on \(f\sigma_8\). Therefore, the same setting is applied to the Fundamental Plane data. Different from Fig.~\ref{fig:mock_TF}, the mean uncertainty (0.064) is close to the standard deviation (0.063) of the best-fit because the Fundamental Plane catalogue has many more galaxies than the Tully-Fisher catalogue. Nonetheless, the mean uncertainty exceeds the standard deviation, indicating that uncertainty is not underestimated. The relative uncertainty on \(f\sigma_8\) is also lower for mocks with higher best-fit \(f\sigma_8\) because they have better signal-to-noise. 
\subsection{Results from testing the combined mocks}
\subsubsection{Only using log-distance ratio}
\begin{figure}
        \centering
	\includegraphics[width=1.0\textwidth]{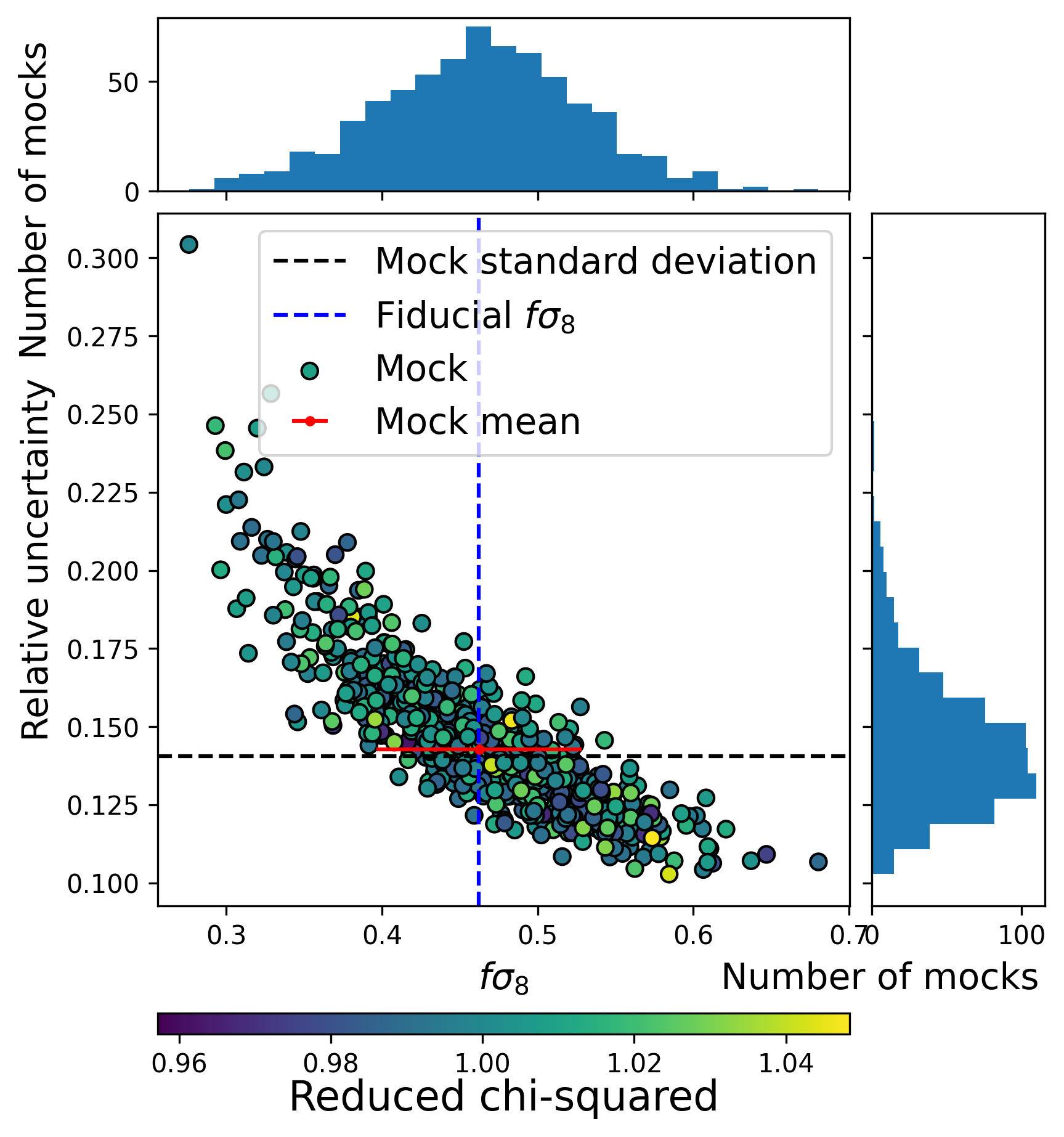}
    \caption{The constraints on \(f\sigma_8\) from 675 combined mocks with \(\sigma_u = 10h^{-1} \mathrm{Mpc}\) and grid size of \(20h^{-1} \mathrm{Mpc}\) using only the log-distance ratio. The colour indicates the reduced chi-squared of each mock. The mock mean is consistent with the true \(f\sigma_8\) indicated by the blue dashed line. }
    \label{fig:mock_combined_pv}
\end{figure}
Fig.~\ref{fig:mock_combined_pv} illustrates the constraints on the growth rate of structure with the combined peculiar velocity catalogue with \(\sigma_u = 10 h^{-1}\mathrm{Mpc}\). The mean \(f\sigma_8=0.462\) is consistent with the fiducial value since both the Tully-Fisher and Fundamental Plane mocks return unbiased \(f\sigma_8\) with \(\sigma_u = 10 h^{-1}\mathrm{Mpc}\). Similar to Fig.~\ref{fig:mock_FP}, the mean uncertainty (0.066) is close to the standard deviation (0.065), indicating that the uncertainty is not underestimated. The reduced chi-squared for all fits is close to 1, indicating that the model fits all mocks well. The relative uncertainty of \(f\sigma_8\) of the combined mocks is similar to that of the Fundamental Plane mocks since over 90\% of galaxies in the combined catalogue are the Fundamental Plane galaxies. The analysis here is a preparation for the DR2 data release, which will contain more Tully-Fisher galaxies, and they will play a more important role in the constraint in the future. 
\subsubsection{Combining log-distance ratio with galaxy overdensity}
\begin{figure}
        \centering
	\includegraphics[width=1.0\textwidth]{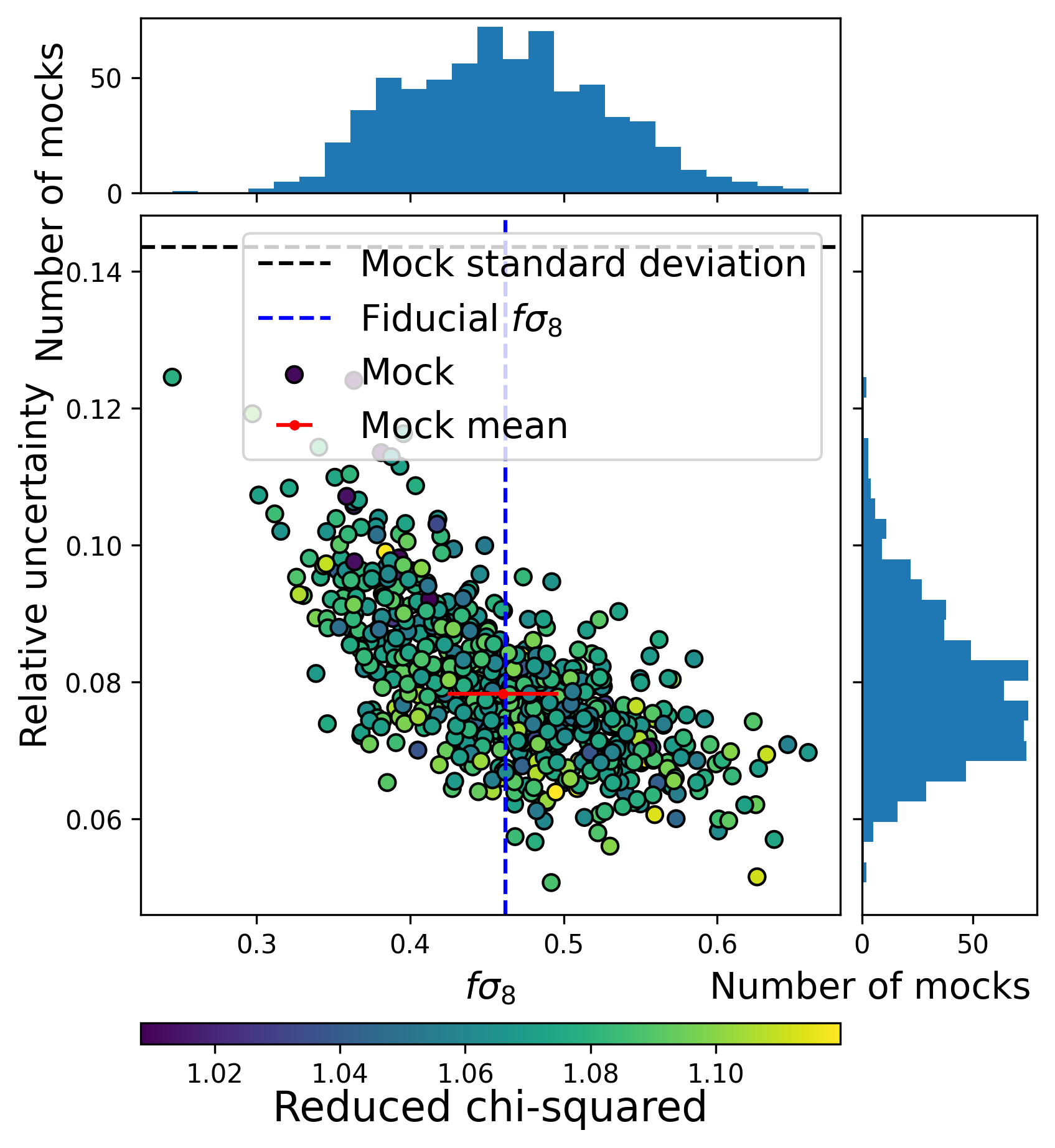}
    \caption{The constraints on \(f\sigma_8\) from 675 combined mocks with \(\sigma_u = 16h^{-1} \mathrm{Mpc}\) and grid size of \(20h^{-1} \mathrm{Mpc}\) using both galaxy overdensity and log-distance ratio. The colour indicates the reduced chi-squared of each mock. The mock mean is consistent with the true \(f\sigma_8\) indicated by the blue dashed line.}
    \label{fig:mock_combined}
\end{figure}
Fig.~\ref{fig:mock_combined} illustrates the constraints on \(f\sigma_8\) from 675 mocks with \(\sigma_u = 16h^{-1} \mathrm{Mpc}\) and grid size of \(20h^{-1} \mathrm{Mpc}\) using both galaxy overdensity and log-distance ratio. The best-fit \(\sigma_u\) is different from that when using only the peculiar velocity catalogue because the cross-covariance also depends on \(\sigma_u\). The mean \(f\sigma_8\) (0.460) constraint is consistent with the fiducial value indicated by the blue dashed line, and the reduced chi-squared is around one. Therefore, we decided to use the same setting when applying to the data catalogue. Different from Fig.~\ref{fig:mock_TF} and~\ref{fig:mock_FP}, the mean uncertainty of \(f\sigma_8\) (0.036) when combining the galaxy overdensity and peculiar velocity is lower than the standard deviation of \(f\sigma_8\) (0.066). There are two possible reasons for this discrepancy. Firstly, the reduced chi-squared for all mocks is close to one, indicating the method is not underestimating the statistical uncertainty. Therefore, the covariance matrix is probably underestimating the sample variance. This is possibly because the gridding kernel significantly reduces the power spectrum and, consequently, the covariance matrix's amplitude (see Fig.\ref{fig:damping_vs_grid_corr}). The gridding does not bias the statistical uncertainty, as indicated by the reduced chi-squared value, since our data vector is also gridded. However, the gridding will remove all small-scale information below 20\(h^{-1}\)Mpc, but the true sample variance includes fluctuations on all scales, not just those larger than the gridding scale. Therefore, the mean uncertainty will be smaller than the standard deviation. Furthermore, equation~(\ref{eq:xi}) is a Fourier transform, so it should be integrated from zero to infinity, not \(k_{\mathrm{min}}\) to \(k_{\mathrm{max}}\). However, we don't have a power-spectrum model for extreme nonlinear scales. This could potentially underestimate the uncertainty. Secondly, we do not expect all mocks to have the same \(\sigma_u\). Therefore, if we let \(\sigma_u\) free, the standard deviation will possibly reduce while the mean uncertainty increases. This will help to reconcile the discrepancy. However, it is not computationally possible to free \(\sigma_u\) with the current method. We decide to follow ref.~\citep{Sugiyama_2023b} to rescale the uncertainty when fitting the data by the ratio of the standard deviation (0.059) and the mean uncertainty (0.036) with the fiducial choices of \(\sigma_u\) and grid size after the redshift and \(4\sigma\) outlier cut. One may notice that after rescaling the uncertainty, the multi-tracer approach yields a similar uncertainty as that with only the peculiar velocity. However, we stress that with only peculiar velocity, \(f\sigma_8\) is very degenerate with \(\sigma_u\) as shown in Fig.~10 of ref.~\citep{Carreres_2023}, so it will have a much larger systematic uncertainty. On the other hand, the galaxy data does not depend on \(\sigma_u\) and could potentially help to break the degeneracy. Additionally, rescaling the uncertainty this way is a conservative approach and may overestimate it. Therefore, if we take \(\sigma_u\) into account, the multi-tracer approach will still give a smaller uncertainty.
\begin{table}[t!]
\centering                          
\renewcommand{\arraystretch}{1.4} 
\begin{tabular}{c|c|c|c|c}        
$\sigma_u$ ($h^{-1}\mathrm{Mpc}$) & mean $f\sigma_8$ & mean uncertainty & standard deviation & grid size \\ \hline \hline
15 & 0.442 & 0.035 (7.9\%) & 0.056 (12.7\%) & 20$h^{-1}\mathrm{Mpc}$ \\
\hline 
16 & 0.460 & 0.036 (7.8\%) & 0.066 (14.3\%) & 20$h^{-1}\mathrm{Mpc}$ \\ 
\hline
17 & 0.478 & 0.038 (8.0\%) & 0.071 (14.9\%) & 20$h^{-1}\mathrm{Mpc}$ \\
\hline
16 & 0.491 & 0.045 (9.2\%) & 0.064 (13.0\%) & 30$h^{-1}\mathrm{Mpc}$ \\
\hline
16 ($4\sigma$ + TF redshift cut) & 0.455 & 0.036 (7.9\%) & 0.059 (13.0\%) & 20$h^{-1}\mathrm{Mpc}$\\
\hline
\end{tabular}
\caption{The mean \(f\sigma_8\) from 675 combined Tully-Fisher and Fundamental Plane mocks with different values of \(\sigma_u\) using both density and velocity fields. Only the statistical uncertainty is included here. The uncertainty of \(f\sigma_8\) is estimated with the Hessian matrix at the maximum likelihood. The mean \(f\sigma_8\) from \(\sigma_u = 16h^{-1}\mathrm{Mpc}\) is closest to the fiducial value (\(f\sigma_8=0.462\)). We also find that the additional quality cuts applied to the data catalogue do not have a strong impact on the best-fit \(f\sigma_8\) in the mocks.}    
\label{tab:mock_sigmau} 
\end{table}

Table~\ref{tab:mock_sigmau} presents the mean \(f\sigma_8\) from 675 mocks with different values of \(\sigma_u\). We cut out all grids with galaxy overdensity \(\delta_g > 10\) because our model is based on the quasi-linear theory, and larger \(\delta_g\) significantly worsens the reduced chi-squared of our fits. Fixing \(\sigma_u = 16 h^{-1} \mathrm{Mpc}\) provides the mean \(f\sigma_8\) closest to the fiducial value of \(0.462\). Additionally, we find that increasing the grid size slightly increases \(f\sigma_8\), which remains consistent with the fiducial value. However, the relative uncertainty is worse for a larger grid size because more small-scale information is removed\footnote{We could not use a smaller grid size than \(20h^{-1}\mathrm{Mpc}\) because firstly, the MCMC analysis will become extremely slow since a smaller grid size will increase the dimension of the covariance matrix and the data vector. Secondly, a smaller grid size will also demand more memory to run the analysis. For \(20h^{-1}\mathrm{Mpc}\) grid size, it takes around \(1\) second per MCMC iteration and 140 Gb of memory.}. Therefore, we decide to use \(20h^{-1} \mathrm{Mpc}\) grid size when fitting the data to obtain a tighter unbiased constraint on \(f\sigma_8\). 

In this work and the companion papers \citep{DESIY1_xi, DESIY1_pk}, we remove Tully-Fisher galaxies beyond the redshift of 0.05 since there are inconsistencies between the constraints on the growth rate with the Tully-Fisher catalogue and the Fundamental Plane catalogue (see appendix \ref{sec:fit_choice} for more detail), possibly due to unknown systematics. Additionally, we apply a \(4\sigma\) cut on the log-distance ratio to remove outliers introduced by systematics. We reran the mocks with these two additional cuts and found that it has little impact on the mean \(f\sigma_8\) and its uncertainty in the mocks. 

Now we have the best-fit \(\sigma_u\), the systematic uncertainty due to fixing \(\sigma_u\) is estimated with the finite difference method \citep{Adams_2020, Lai_2022}
\begin{equation}
    \sigma_{\sigma_u} = \frac{\partial f\sigma_8}{\partial \sigma_u} \approx \frac{f\sigma_8(\sigma_u + \delta_{\sigma_u}) - f\sigma_8(\sigma_u - \delta_{\sigma_u})}{2\delta_{\sigma_u}} = 0.018. 
    \label{eq:sigma_sys}
\end{equation}
This will be added to the statistical uncertainty in quadrature to give the final uncertainty on \(f\sigma_8\). The mean relative uncertainty and the standard deviation are roughly constant when choosing different values of \(\sigma_u\) as expected, consistent with ref.~\citep{Lai_2022}. 

\section{Data fitting and discussion}
\label{sec:result}

\begin{figure}
    \centering
	\includegraphics[width=0.7\textwidth]{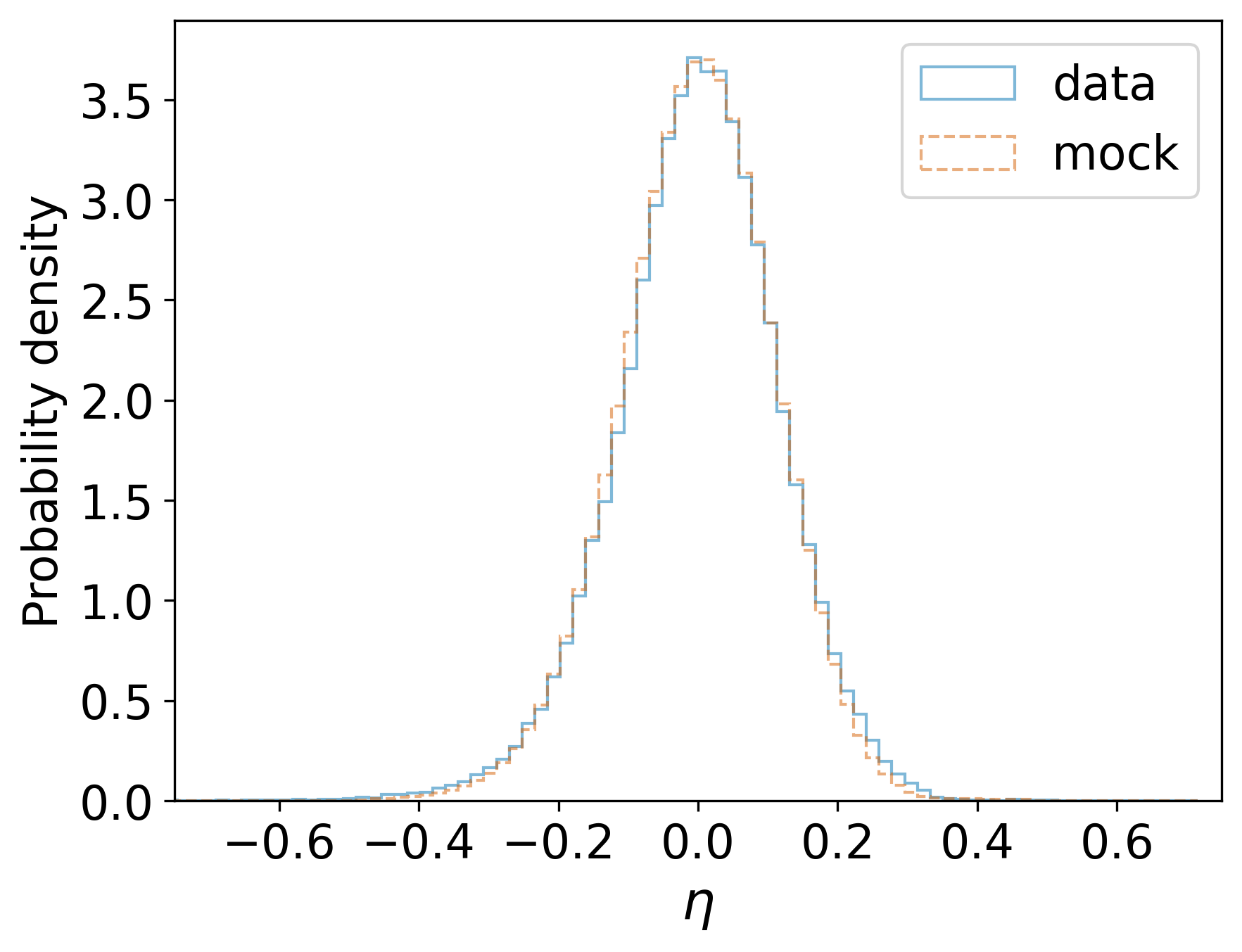}
    \caption{The probability density distribution of log-distance ratio in the mocks (dashed orange) and the data (solid blue). We combine log-distance ratio measurements in all 675 mocks and then normalise both the mocks and data to generate this plot. It shows that the distribution of log-distance ratio in the mocks is consistent with the data.}
    \label{fig:mock_vs_data_hist}
\end{figure}

Before applying our method to the data, we want to ensure that the mocks accurately replicate the distribution of the log-distance ratio in the data. Fig.~\ref{fig:mock_vs_data_hist} illustrates the probability density distribution of log-distance ratio in the mocks and data after applying the \(4\sigma\) cut to remove the outliers.\footnote{Theoretically, the log-distance ratio should be Gaussian distributed. Based on our sample size, there should be around \(\sim 10\) peculiar velocity measurements that are more than \(4\sigma\) from the median. However, we remove on the order of \(100\) galaxies after applying the \(4\sigma\) cut. This indicates that most of the removed galaxies are statistical outliers.} After this cut, the final number of galaxies in the DESI DR1 peculiar velocity sample is \(76616\)\footnote{In the companion paper, ref.~\citep{DESIY1_xi}, their final sample has 76615 galaxies. This is because they applied the \(4\sigma\) cut to the Fundamental Plane and Tully-Fisher catalogue separately. However, we do not expect the extra galaxy to have a significant impact on our constraint and conclusion in this work.} with an effective redshift of \(z_{\mathrm{eff}} = 0.07\).

\begin{figure}
        \centering
	\includegraphics[width=1.0\textwidth]{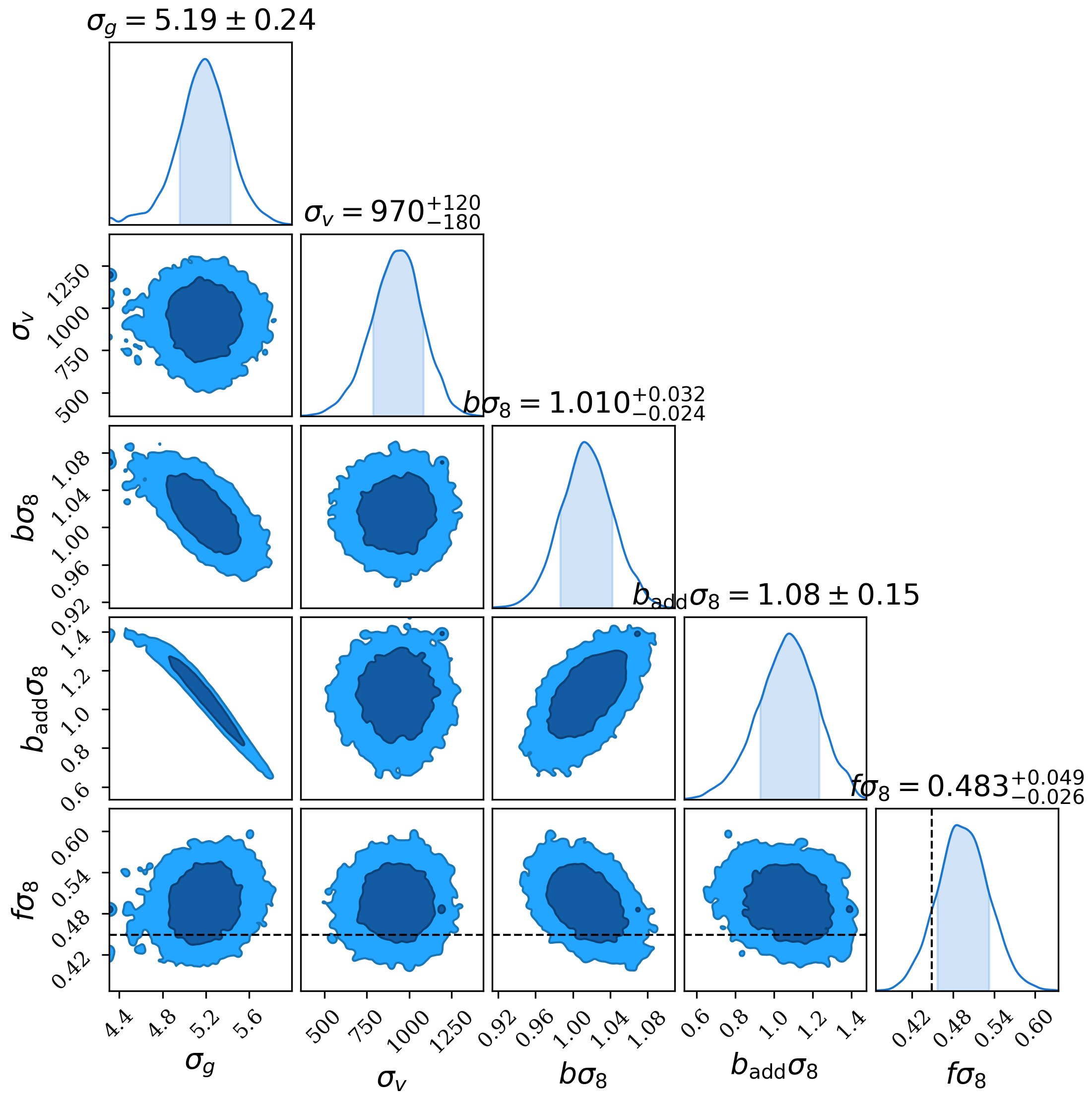}
    \caption{The constraints on \(f\sigma_8\) with the BGS galaxy, Fundamental Plane, and Tully-Fisher peculiar velocity catalogues. The uncertainty in \(f\sigma_8\) here does not include the systematic uncertainty. We follow the companion papers \citep{DESIY1_xi, DESIY1_pk} and exclude all Tully-Fisher log-distance ratio measurements beyond a redshift of 0.05. Additionally, we apply a galaxy overdensity cut, \(\delta_g^{\mathrm{cut}} = 10\), to remove grids that our quasi-linear theory cannot accurately model, and a cut on the log-distance outliers, \(\sigma_{\mathrm{cut}} = 4\). Our constraint on \(f\sigma_8\) is slightly higher than the GR prediction (indicated by the black dashed line, \(f\sigma_8 = 0.444\)) at around a \(1.3\sigma\) level. After accounting for the systematic uncertainty of fixing \(\sigma_u\), the discrepancy is reduced to around \(1\sigma\) level.}
    \label{fig:data_combined}
\end{figure}

In Section~\ref{sec:mock_test}, we show that changing fitting choices such as grid size and \(4\sigma\) cut modify the constraint on \(f\sigma_8\) less than its uncertainty in the mocks. In appendix~\ref{sec:fit_choice}, we demonstrate that the same is true for the data catalogue. Therefore, in this section, we will constrain \(f\sigma_8\) with our fiducial fitting choices. Fig.~\ref{fig:data_combined} illustrates the statistical constraints with the BGS galaxy, Fundamental Plane, and Tully-Fisher peculiar velocity catalogue. We apply a galaxy overdensity cut \(\delta_g^{\mathrm{cut}} = 10\) to remove grids that our quasi-linear theory cannot accurately model. In appendix \ref{sec:fit_choice}, we demonstrate that these cuts do not impact the constraints. Our constraint is \(f\sigma_8 = 0.483_{-0.026}^{+0.049}(\mathrm{stat}) \pm 0.018(\mathrm{sys})\). The relative statistical uncertainty on \(f\sigma_8\) is \(7.8\%\), consistent with the relative mean uncertainty from the mocks in table~\ref{tab:mock_sigmau}. However, during the mock test, we noticed that the mean uncertainty is lower than the standard deviation. To fully account for the sample variance, we inflate the error bar by \(\frac{0.059}{0.036}\approx1.64\)\footnote{0.059 is the mock standard deviation, and 0.036 is the mock mean uncertainty in table~\ref{tab:mock_sigmau}. This will rescale the under-estimated uncertainty to be consistent with the mock standard deviation.}. After inflating the uncertainty, the final constraint is \(f\sigma_8 = 0.483_{-0.043}^{+0.080}(\mathrm{stat}) \pm 0.018(\mathrm{sys})\). The relative uncertainty increases to \(13\%\) after including the systematic uncertainty. It is about 30\% tighter than the constraint from the SDSS peculiar velocity catalogue \citep{Lai_2022} because the DESI sample contains more galaxy overdensity and log-distance ratio measurements. The \(f\sigma_8\) mean of the posterior is higher than the prediction from GR with the Planck \(\Lambda\)CDM cosmology (\(f\sigma_8 = 0.449\)), indicated by the black dashed line. However, after accounting for the systematic uncertainty of fixing \(\sigma_u\), our constraint is consistent with the GR prediction at around \(1\sigma\) level. Our constraint on \(\sigma_g\) is consistent with the values in ref.~\citep{Koda_2014} from simulations. Similar to ref.~\citep{Lai_2022}, \(\sigma_g\) is strongly degenerate with \(b_{\mathrm{add}}\sigma_8\) since both changes the amplitude of the covariance matrix. We will leave for future work to investigate whether it will be possible to eliminate the \(b_{\mathrm{add}}\sigma_8\) parameter with a better small-scale modelling of the power spectrum. The constraint on \(f\sigma_8\) is also consistent with the momentum power spectrum and the velocity correlation function approaches in the companion papers \citep{DESIY1_xi, DESIY1_pk}. However, \(\sigma_v\) is around \(3\sigma\) higher than previous constraints, which are usually around \(300 \mathrm{km}\mathrm{s}^{-1}\) \citep{Howlett_2017, Adams_2017, Adams_2020, Lai_2022}. This result is consistent with best-fits from the Fundamental Plane mocks, which are around \(\sim 1000\mathrm{km}\mathrm{s}^{-1}\), so this is not likely to be due to systematics in the data catalogue. There are two possible reasons behind the high \(\sigma_v\) value. Firstly, we are using a \(20 h^{-1} \mathrm{Mpc}\) grid size, the minimum scale that can be probed with this grid size is approximately \(k \sim \frac{2\pi}{r} \sim 0.3h\mathrm{Mpc}^{-1}\), the quasi-linear theory here breaks down on such small scale. Since \(\sigma_v\) is the only nonlinear velocity parameter, it not only accounts for the nonlinear dispersion, but also other nonlinear effects that are not taken into account by our model here. Secondly, Fig.~\ref{fig:damping_vs_grid_corr} demonstrates that the grid correction function is still around 0.2 beyond \(k = 0.20h\mathrm{Mpc}^{-1}\) with a \(20h^{-1}\mathrm{Mpc}\) grid size. There is still a contribution from smaller scales to the covariance matrix. For the galaxy auto-covariance matrix, the \(b_{\mathrm{add}}\sigma_8\) parameter accounts for these contributions, but for the velocity auto-covariance matrix, the only term that is added to it is \(\sigma_v\), so a higher \(\sigma_v\) also accounts for the contribution to the covariance matrix on smaller scales. Nonetheless, the reduced chi-squared of our fit is around 1.011 with 10,716 degrees of freedom, indicating our model is a good fit for the data.

\subsection{Comparison with other DESI PV measurements}
\begin{figure}
        \centering
	\includegraphics[width=1.0\textwidth]{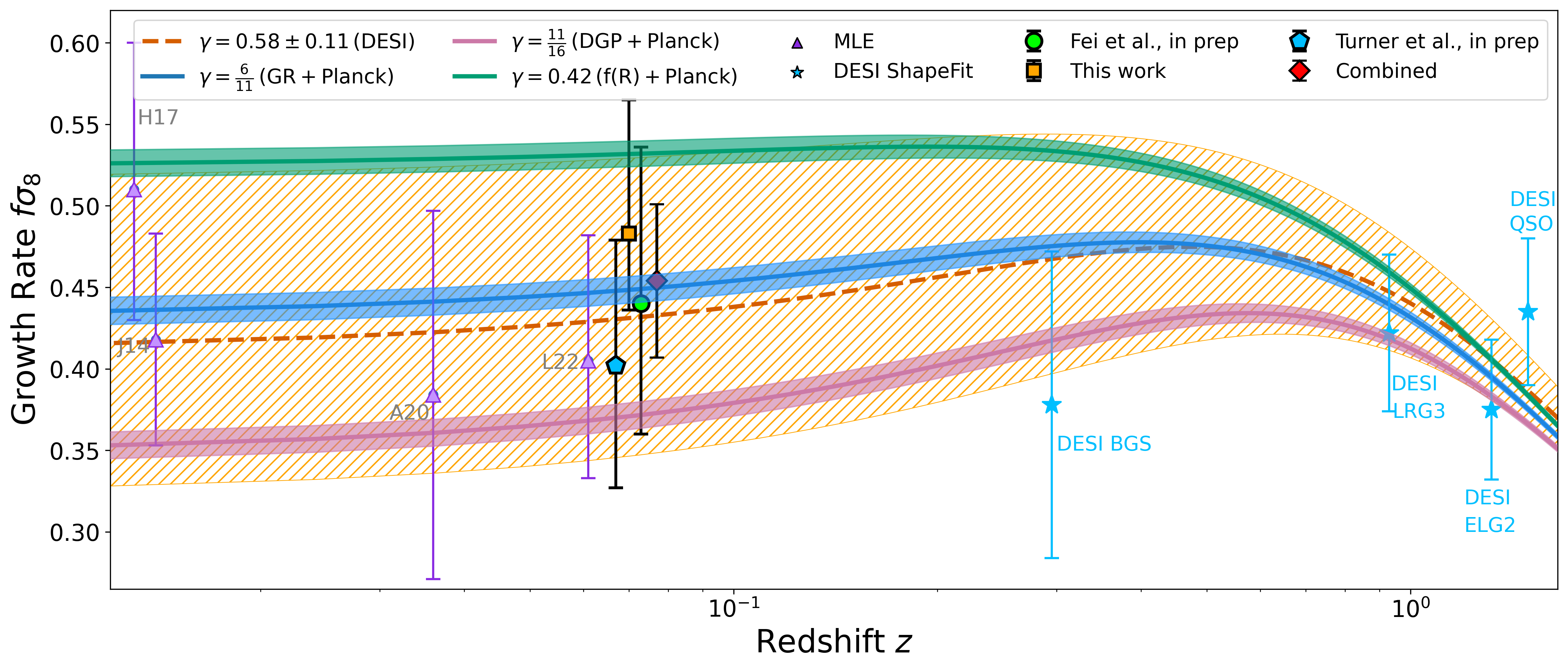}
    \caption{Growth rate of structure measurements across different redshifts. The three measurements from DESI are shown by the green circle (momentum power spectrum), the yellow square (maximum likelihood fields method, this work), and the blue pentagon (velocity correlation function). The red diamond shows the combined constraint. All four measurements have the same effective redshift but are offset slightly to differentiate them. The blue band shows the GR prediction with the Planck \(\Lambda\)CDM cosmology. The red and green bands show the predictions from DGP and f(R) gravity models \citep{Linder_2007} within the Planck cosmology, respectively. Lastly, the orange band indicates the best-fit \(\gamma\) by combining the combined constraint from peculiar velocity (red diamond) with the \textit{ShapeFit} constraints on \(f\sigma_8\) in higher redshifts (blue stars) \citep{Adame_2025c}. The purple triangles at low redshift correspond to previous measurements using the maximum likelihood fields method \citep{Howlett_2017, Johnson_2014, Adams_2020, Lai_2022}.}
    \label{fig:gamma_plot}
\end{figure}

Besides the maximum likelihood fields method, DESI also use the velocity correlation function and momentum power spectrum to measure the growth rate of structure. The velocity correlation function companion paper \citep{DESIY1_xi} shows that the constraint is \(f\sigma_8 = 0.402_{-0.077}^{+0.075}\) and the momentum power spectrum companion paper \citep{DESIY1_pk} shows that the constraint is \(f\sigma_8 = 0.443^{+0.110}_{-0.070}\). All three measurements are consistent within \(1\sigma\) with each other and also consistent with the Planck\(+\Lambda\)CDM prediction. Fig.~\ref{fig:gamma_plot} summarises the measurements from the DESI DR1 peculiar velocity catalogue and other previous constraints on the growth rate of structure with the maximum likelihood fields method \citep{Howlett_2017, Johnson_2014, Adams_2020, Lai_2022} and DESI \textit{ShapeFit} constraints on \(f\sigma_8\) from higher redshifts \citep{Adame_2025c}. The best-fit growth index is consistent with the GR prediction and the Planck \(\Lambda\)CDM cosmology within the uncertainty. Furthermore, Fig.~11 in ref.~\citep{DESIY1_pk} illustrates that combining DESI with other datasets at low redshift could significantly improve the constraint on the growth index, potentially reaching a similar relative uncertainty as Planck. 

Despite consistent \(f\sigma_8\) constraints from all three methods, the correlation function and the power spectrum have around 30\% larger relative uncertainty of \(f\sigma_8\) than that from the maximum likelihood fields method. To understand this difference, we first compare the constraints on the growth rate from these three approaches with the mocks. The mean relative uncertainty for the maximum likelihood method is \(13\%\) (after including the systematic uncertainty and inflating the statistical uncertainty to be equal to the systematic uncertainty), \(11\%\) for the momentum power spectrum (Fig.~6 in the companion momentum power spectrum paper \citep{DESIY1_pk}), and \(16\%\) for the correlation function (Fig.~6 in the companion velocity correlation function paper \citep{DESIY1_xi}). The maximum likelihood method is roughly consistent with the momentum power spectrum when fitting the mocks, while the correlation function has a slightly higher relative uncertainty. Both the maximum likelihood method and the momentum power spectrum probes to \(\sim0.3 h\mathrm{Mpc}^{-1}\). The fitting range for the correlation function is \(24 h^{-1}\mathrm{Mpc} - 120 h^{-1}\mathrm{Mpc}\). Using the \(k\sim \frac{2\pi}{r}\) approximation, this corresponds to \(0.05h^{-1}\mathrm{Mpc} - 0.26^{-1}\mathrm{Mpc}\). Both the momentum power spectrum and the maximum likelihood fields method probe larger scales than the correlation function. Since the velocity field signal is mostly on large scales, both the momentum power spectrum and the maximum-likelihood fields method yield tighter constraints on \(f\sigma_8\) than the velocity correlation function. Fig.~7 in the companion velocity correlation function paper \citep{DESIY1_xi} also demonstrates that the relative uncertainty from the correlation function approach will reduce when including larger scales. 

However, when fitting to the data, we only fit the momentum power spectrum up to \(k_{\mathrm{max}} = 0.1h\mathrm{Mpc}^{-1}\) \citep{DESIY1_pk}. While the velocity signal is mainly on large scales, it seems like small-scale information is vital for the momentum power spectrum method to break the degeneracy between two nonlinear velocity dispersion parameters with \(f\sigma_8\) (Fig.~5 and Fig.~9 in the momentum power spectrum companion paper \citep{DESIY1_pk}). These constraints on these nonlinear dispersion parameters are weaker because of lower \(k_{\mathrm{max}}\), so the \(f\sigma_8\) constraint is also wider. On the other hand, both Fig.~\ref{fig:mock_combined} in this work and Fig.~5 in the velocity correlation function companion paper \citep{DESIY1_xi} demonstrate that the relative uncertainty is inversely proportional to the best-fit \(f\sigma_8\) due to higher signal-to-noise for higher \(f\sigma_8\). The best-fit \(f\sigma_8\) for the correlation function is lower for the data than the mocks, so its relative uncertainty also increases slightly from 16\% to 19\%. In contrast, the best-fit \(f\sigma_8\) from the maximum likelihood method is roughly the same between the data and the mocks, keeping the relative uncertainty at around 13\%. Consequently, the growth rate of structure constraint from the maximum likelihood fields method is approximately 30\% tighter than that from the velocity correlation function or the momentum power spectrum.

Since the three methods are not in statistical tension with each other, we produce a combined constraint on \(f\sigma_8\) with the approach in ref.~\citep{Sanchez_2016}. We compute the correlation between different methods with our growth rate constraints from the same mocks. We refer readers to the velocity correlation function companion paper \citep {DESIY1_xi} for a detailed discussion of the process of combining the three measurements by applying the method in ref.~\citep{Sanchez_2016}. The final consensus constraint on the growth rate of structure is \(f\sigma_8 (z_{\mathrm{eff}}=0.07) = 0.450\pm0.055\), consistent with GR prediction with Planck \(\Lambda\)CDM cosmology within \(1\sigma\). Combining with the high redshift growth rate of structure measurements from DESI \textit{ShapeFit}, we find the growth index is \(\gamma = 0.58\pm0.11\) \citep{DESIY1_pk}. We refer the readers to the momentum power spectrum companion paper \citep{DESIY1_pk} for more details on constraining \(\gamma\).     

\section{Conclusion}
\label{sec:conclusion}
We present constraints on the growth rate of structure by combining the DESI DR1 BGS sample, the Fundamental Plane, and the Tully-Fisher peculiar velocity catalogue using the maximum likelihood fields method. The combined DESI DR1 peculiar velocity catalogue contains approximately ten times as many galaxies and two times as many log-distance ratio measurements as the largest existing single peculiar velocity catalogue, the SDSS peculiar velocity catalogue. To speed up the analysis, we rewrite the code with \textsc{JAX}. This means we no longer need to approximate the likelihood function using its second-order Taylor expansion \citep{Lai_2022}, thereby improving the accuracy of our constraints. Our tests on the mock catalogues demonstrate that our model can return unbiased constraints on the growth rate of structure. We find that increasing the grid size does not affect the constraint on \(f\sigma_8\) with the log-distance ratio, consistent with ref.~\citep{Howlett_2017}, because peculiar velocities are most sensitive to large scales. However, increasing the grid size removes the small-scale information and significantly weakens the constraint on \(f\sigma_8\) when using only galaxy overdensity. After accounting for the systematic bias of fixing \(\sigma_u = 16h^{-1} \mathrm{Mpc}\), removing outliers from the data catalogue, and inflating the uncertainty to match the spread of \(f\sigma_8\) in the mocks, our constraint is \(f\sigma_8 = 0.483_{-0.043}^{+0.080}(\mathrm{stat}) \pm 0.018(\mathrm{sys})\) with a \(20h^{-1}\mathrm{Mpc}\) grid size, consistent with the GR prediction at around \(1\sigma\) level. The reduced chi-squared value is approximately 1.011, indicating that our model is a good fit to the data catalogue. Our measurement is around 30\% tighter than that from the SDSS peculiar velocity catalogue. In companion papers \citep{DESIY1_pk, DESIY1_xi}, we present complementary results using the same data but with different analysis methods: the velocity correlation function and the momentum power spectrum. We find all three measurements are consistent with each other within \(1\sigma\). After combining the three measurements, our consensus constraint on the growth rate is \(f\sigma_8 (z_{\mathrm{eff}}=0.07) = 0.450\pm0.055\), consistent with the GR prediction with the Planck \(\Lambda\)CDM cosmology (0.449) with \(1\sigma\). Lastly, using the combined constraint on the growth rate and \textit{ShapeFit} constraints on the growth rate from DESI at higher redshifts, ref.~\citep{DESIY1_pk} demonstrate that the constraint on the growth index from DESI alone is \(\gamma = 0.58\pm0.11\). 

The systematic uncertainty of fixing the nonlinear RSD damping parameters \(\sigma_u\) is around 30\% of the statistical uncertainty. However, future DESI data releases will provide additional galaxy and peculiar velocity measurements, and the systematic uncertainty could potentially exceed the statistical uncertainty. To remove the systematic uncertainty, we need to pre-compute covariance matrices with different \(\sigma_u\) values and use the linear interpolation during the MCMC. However, since the covariance matrices are large, fixing \(\sigma_u\) already requires around \(\sim 140\) GB of memory to perform the analysis. This is not currently feasible with a \(20h^{-1}\mathrm{Mpc}\) grid size. Increasing the grid size could reduce the memory required for the analysis, but it will also weaken the constraint on \(f\sigma_8\). However, lossless data compression techniques such as MOPED\footnote{Massively Optimised Parameter Estimation and Data compression} \citep{Heavens_2000, Heavens_2017, Heavens_2023, Alsing_2018, Lai_2024} could massively reduce the dimension of the covariance matrix and data without weakening the constraints. Future research could focus on applying data compression to the maximum likelihood fields method to reduce the dimension of the covariance matrices, thereby allowing us to include \(\sigma_u\) as a free parameter and properly treat this contaminant. 

\appendix
\section{Comparing \(f\sigma_8\) constraints with SPT and RPT}
\label{sec:SPT_vs_RPT}
In this work, we generate power spectra from SPT (Standard Perturbation Theory) rather than RPT (Renormalised Perturbation Theory), ensuring that all three pipelines share a consistent theoretical basis. In this section, we compare the power spectra from SPT and RPT with the baseline \(\Lambda\)CDM cosmology from \textsc{AbacusSummit}. 

\begin{figure}
        \centering
	\includegraphics[width=0.495\textwidth]{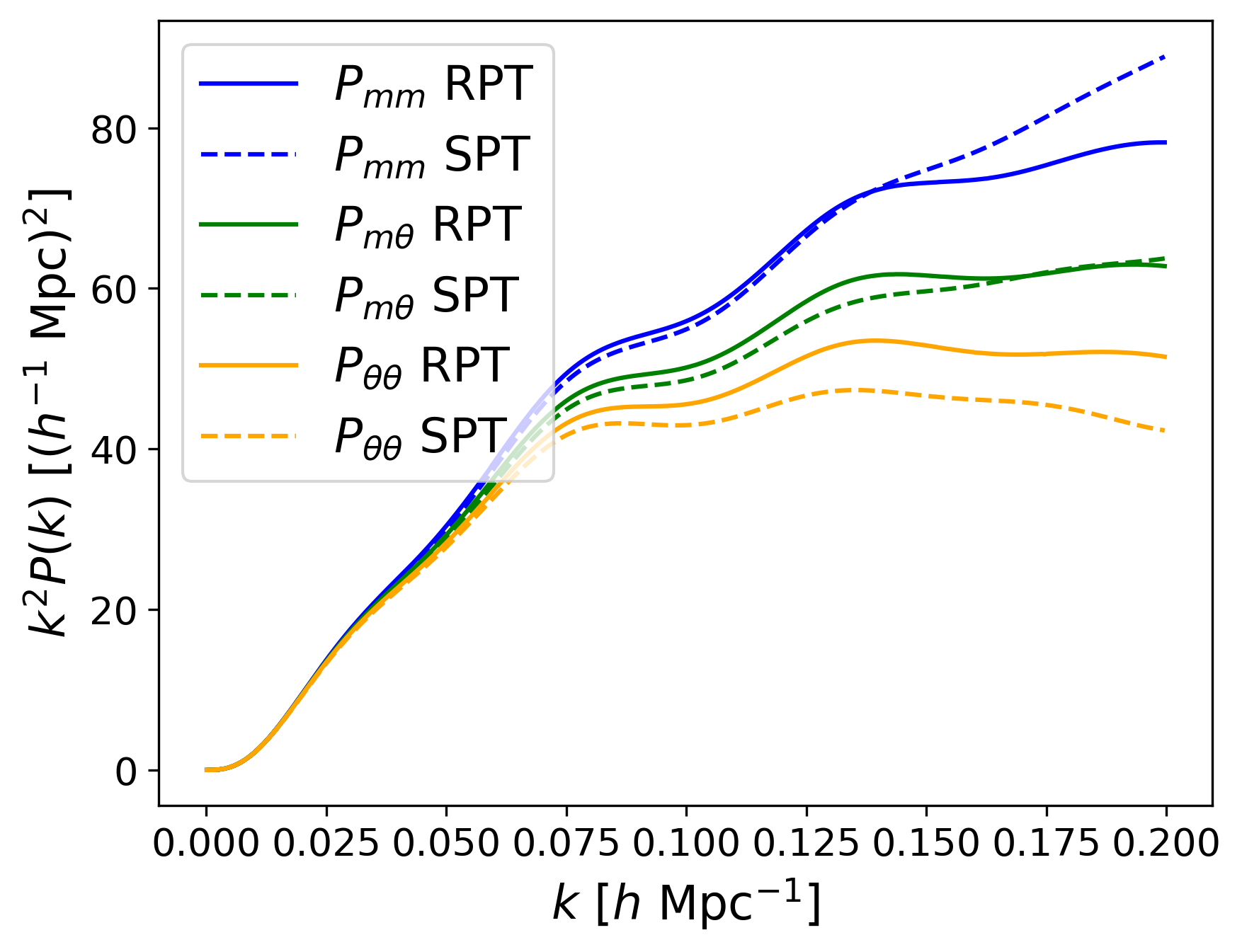}
        \includegraphics[width=0.495\textwidth]{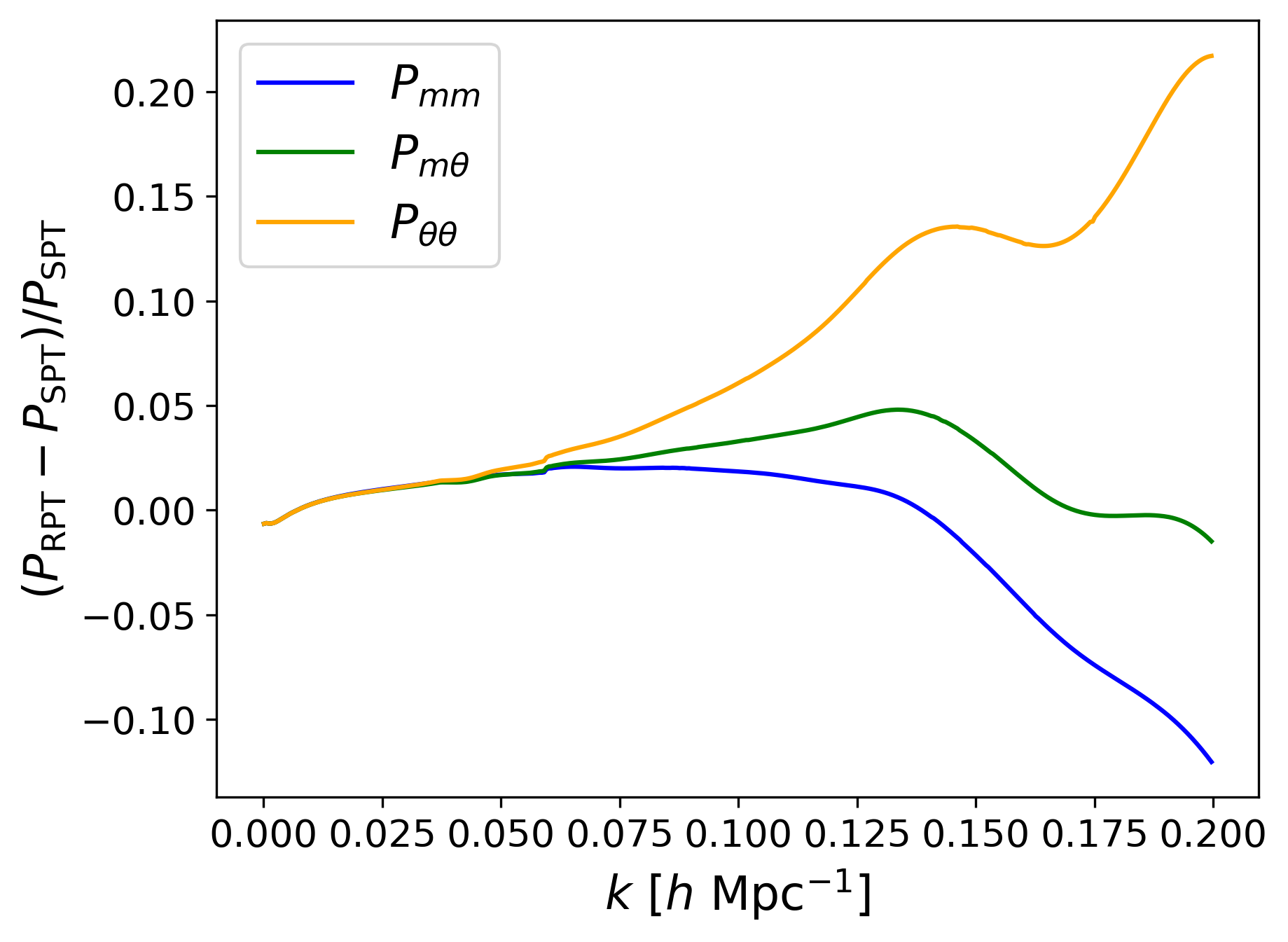}
    \caption{Left: matter auto-power spectrum, matter-velocity divergence cross-power spectrum, and the velocity divergence auto-power spectrum from SPT and RPT. Right: The fractional difference between the RPT and SPT power spectra. At large scales, both models return consistent power spectra because they reduce to linear perturbation theory. The difference between the two models reaches \(\sim20\%\) for the velocity divergence auto-power spectrum, \(\sim 10\%\) for the matter auto-power spectrum, and \(\sim 5\%\) for the cross-power spectrum. }
    \label{fig:SPT_vs_RPT}
\end{figure}

\begin{figure}
    \centering
    \includegraphics[width=1.0\linewidth]{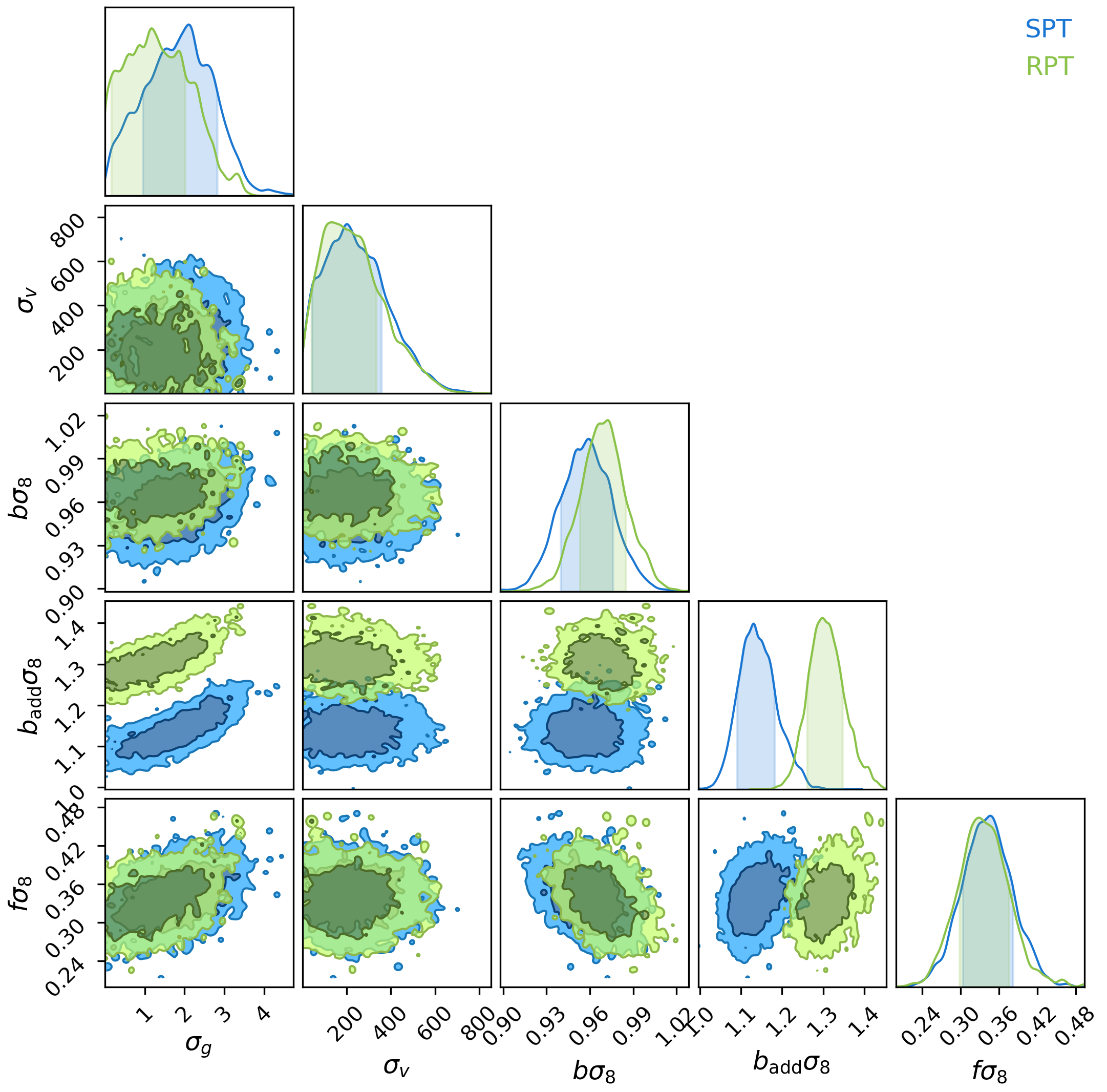}
    \caption{Constraints on free parameters with the SPT power spectra and the RPT power spectra for mock 0. We set \(\delta_g^{\mathrm{cut}} = 10\), \(\sigma_{\mathrm{cut}} = 4\), \(\sigma_u = 16 h^{-1}\mathrm{Mpc}\), \(20h^{-1}\mathrm{Mpc}\) grid size, and no redshift cut. Both models give consistent constraints on \(f\sigma_8\). The differences between the two models shown in Fig.~\ref{fig:SPT_vs_RPT} are mitigated by the nuisance parameters.}
    \label{fig:SPT_RPT_MCMC}
\end{figure}

Fig.~\ref{fig:SPT_vs_RPT} compares power spectra from SPT and RFT. At large scales (\(\lesssim 0.05 h \mathrm{Mpc}^{-1}\)), both models produce consistent power spectra since they both reduce to the linear perturbation theory on large scales. At small scales, the differences between the two models reach \(\sim20\%\) for the velocity divergence auto-power spectrum, \(\sim 10\%\) for the matter auto-power spectrum, and \(\sim 5\%\) for the cross-power spectrum. This is because while SPT continues treating the small-scale perturbation perturbatively, while RPT introduces a nonlinear propagator to suppress the fluctuation on small scales \citep{Crocce_2006a}. Fig.~\ref{fig:SPT_RPT_MCMC} illustrates that both SPT and RPT approaches return the same constraint on \(f\sigma_8\) for mock 0. The nuisance parameters mitigate the differences between the two models on small scales.  

\section{The effect of fitting choices on the \(f\sigma_8\) constraint}
\label{sec:fit_choice}

\begin{table}[t!]
\centering                          
\renewcommand{\arraystretch}{1.4} 
\begin{tabular}{c|c|c|c|c|c|c|c}        
Data & $f\sigma_8$ & Reduced $\chi^2$ & dof & grid size & $\sigma_u$ & $\sigma_{\mathrm{cut}}$ & $z_{\mathrm{cut}}$\\ \hline \hline
TF  &$0.85_{-0.11}^{+0.11}$ & $1.08$ & $1653$ & $20$ &$10$ &$3$ & None\\
\hline
TF  &$0.92_{-0.12}^{+0.06}$ & $1.10$ & $1657$ & $20$ &$10$ &$4$ & None\\
\hline
TF  &$0.76_{-0.06}^{+0.14}$ & $1.08$ & $1653$ & $20$ &$0$ &$3$ & None\\
\hline
TF  &$0.79_{-0.08}^{+0.13}$ & $1.08$ & $829$ & $30$ &$10$ &$3$ & None\\
\hline
\hline
TF  &$0.64_{-0.21}^{+0.12}$ & 1.05 & $446$ & $20$ &$10$ &$3$ & $0.05$\\
\hline
TF  &$0.60_{-0.13}^{+0.20}$ & 1.08 & $447$ & $20$ &$10$ &$4$ & $0.05$\\
\hline
TF  &$0.62_{-0.18}^{+0.11}$ & 1.06 & $446$ & $20$ &$0$ &$3$ & $0.05$\\
\hline
TF  &$0.51_{-0.10}^{+0.21}$ & 1.06 & $208$ & $30$ &$10$ &$3$ & $0.05$\\
\hline
\end{tabular}
\caption{Constraints on \(f\sigma_8\) with the Tully-Fisher catalogue with only the log-distance ratio. The full Tully-Fisher sample produces significantly higher \(f\sigma_8\) than that from the Fundamental Plane catalogue in table~\ref{tab:FP_data}. Reducing \(\sigma_u\) and applying a \(3\sigma\) cut to the data will slightly reduce \(f\sigma_8\), but these changes are much smaller than the uncertainty. This result is consistent with the companion papers \citep{DESIY1_pk, DESIY1_xi}, which also find the constraint on growth rate from the Tully-Fisher catalogue inconsistent with that from the Fundamental Plane catalogue. The underlying cause of the inconsistency is unknown, so we follow our companion papers \citep{DESIY1_pk, DESIY1_xi} to remove all Tully-Fisher log-distance ratios beyond the redshift of 0.05 to obtain a conservative constraint on \(f\sigma_8\). This reduces \(f\sigma_8\), and the resultant measurement is consistent with the constraints from the Fundamental Plane catalogue within \(1\sigma\), independent of grid size, \(\sigma_u\), and $\sigma_{\mathrm{cut}}$. }    
\label{tab:TF_data} 
\end{table}

Table~\ref{tab:TF_data} presents the constraint on \(f\sigma_8\) with the Tully-Fisher data catalogue with only the log-distance ratio. Despite the reduced chi-squared indicating that our method is a good fit to the data, using the full Tully-Fisher sample yields \(f\sigma_8\) constraints that are significantly higher than those from the Fundamental Plane catalogue in table~\ref{tab:FP_data}. This result is consistent with the constraints from the velocity correlation function approach in the companion papers \citep{DESIY1_xi}, which find excess correlation between peculiar velocity measurements from the Tully-Fisher catalogue beyond the redshift of 0.05 compared to the Fundamental Plane catalogue. Increasing the grid size and applying a \(3\sigma\) cut\footnote{A \(3\sigma\) cut is likely more appropriate for the Tully-Fisher catalogue, since it contains only 6806 galaxies, less than one-tenth of the combined catalogue. A \(4\sigma\) cut is not enough to cut out all outliers since the number of galaxies is low.} or reducing \(\sigma_u\) can slightly reduce the mean \(f\sigma_8\), but the changes are much smaller than the uncertainty. To reach such a high value of \(f\sigma_8\) at low redshift, either \(\sigma_8\) is two times higher than the measurements from weak lensing \citep{Abbott_2022} and Planck \citep{Planck_2020} or the growth index \(\gamma \sim 0.05\), much stronger than other constraints on \(\gamma\) and theoretical predictions for gravity, including modified gravity models. The source of this inconsistency is unknown, but this is the first time the Tully-Fisher relation has been used to measure peculiar velocity beyond the redshift of 0.05. This could be caused by previously unknown systematics at high redshift. Therefore, we decided to remove all log-distance ratio measurements beyond redshift 0.05 from the Tully-Fisher catalogue to obtain a conservative constraint on \(f\sigma_8\). This reduces the \(f\sigma_8\) constraint from the Tully-Fisher catalogue to be consistent with that from the Fundamental Plane catalogue within \(1\sigma\). This result is also independent of the grid size, \(\sigma_{\mathrm{cut}}\), and \(\sigma_u\). We will investigate this systematics in the Tully-Fisher catalogues in more detail in the future.  

\begin{table}[t!]
\centering                          
\renewcommand{\arraystretch}{1.4} 
\begin{tabular}{c|c|c|c|c|c|c|c}        
Data & $f\sigma_8$ & Reduced $\chi^2$ & dof & grid size & $\sigma_u$ & $\sigma_{\mathrm{cut}}$ & $z_{\mathrm{cut}}$\\ \hline \hline
FP  &$0.510_{-0.035}^{+0.077}$ & $1.00$ & $4779$ & $20$ &$10$ &$4$ & None\\
\hline
FP  &$0.520_{-0.054}^{+0.054}$ & $0.97$ & $4775$ & $20$ &$10$ &$3$ & None\\
\hline
FP  &$0.520_{-0.060}^{+0.060}$ & $1.00$ & $4779$ & $20$ &$0$ &$4$ & None\\
\hline
FP  &$0.512_{-0.061}^{+0.057}$ & $1.05$ & $1784$ & $30$ &$10$ &$4$ & None\\
\hline
\end{tabular}
\caption{Constraints on \(f\sigma_8\) with the Fundamental Plane catalogue with only the log-distance ratio. Increasing grid size, reducing \(\sigma_u\), and applying a \(3\sigma\) cut can sightly reduce \(f\sigma_8\), consistent with table~\ref{tab:TF_data}. However, these reductions in \(f\sigma_8\) are small when compared to the uncertainty of the measurement. Our constraints on \(f\sigma_8\) are generally consistent with the mock mean and the GR prediction within \(1\sigma\).}    
\label{tab:FP_data} 
\end{table}

We then apply our pipeline to the Fundamental Plane catalogue, and the results are shown in table~\ref{tab:FP_data}. Our constraints are consistent with the mock mean and the prediction from GR within \(1\sigma\) independent of \(\sigma_u\), grid size, and \(\sigma_{\mathrm{cut}}\). Although changing \(\sigma_u\), \(\sigma_{\mathrm{cut}}\), and grid size also change the \(f\sigma_8\) mean of the posterior, these changes are small compared to the uncertainty. Similar to the Tully-Fisher catalogue, the reduced chi-squared of our constraints from the Fundamental Plane catalogue is also close to one, indicating our model is a good fit to the data. 

\begin{table}[t!]
\centering                          
\renewcommand{\arraystretch}{1.4} 
\begin{tabular}{c|c|c|c|c|c|c|c}        
Data & $f\sigma_8$ & Reduced $\chi^2$ & dof & grid size & $\sigma_u$ & $\sigma_{\mathrm{cut}}$ & $z_{\mathrm{cut}}$\\ \hline \hline
TF + FP &$0.559_{-0.051}^{+0.053}$ & $0.99$ & $4798$ & $20$ &$10$ &$4$ & None\\
\hline
TF + FP &$0.501_{-0.042}^{+0.057}$ & $0.97$ & $4798$ & $20$ &$10$ &$3$ & None\\
\hline
TF + FP &$0.530_{-0.037}^{+0.058}$ & $1.00$ & $4811$ & $20$ &$0$ &$4$ & None\\
\hline
TF + FP &$0.533_{-0.051}^{+0.048}$ & $1.06$ & $1786$ & $30$ &$10$ &$4$ & None\\
\hline
\hline
TF + FP &$0.532_{-0.067}^{+0.041}$ & $1.00$ & $4787$ & $20$ &$10$ &$4$ & 0.05 TF cut\\
\hline
TF + FP &$0.514_{-0.067}^{+0.067}$ & $0.97$ & $4785$ & $20$ &$10$ &$3$ & 0.05 TF cut\\
\hline
TF + FP &$0.493_{-0.036}^{+0.071}$ & $1.00$ & $4787$ & $20$ &$0$ &$4$ & 0.05 TF cut\\
\hline
TF + FP &$0.513_{-0.061}^{+0.043}$ & $1.06$ & $1784$ & $30$ &$10$ &$4$ & 0.05 TF cut\\
\hline

\end{tabular}

\caption{Constraints on \(f\sigma_8\) with the combined catalogue with only the log-distance ratio. Our constraints are consistent with the results from the Fundamental Plane catalogue, since over \(90\%\) of the galaxies in the combined catalogue are Fundamental Plane galaxies. Applying the redshift cut on the Tully-Fisher catalogue slightly reduces \(f\sigma_8\) since Table~\ref {tab:TF_data} demonstrates that the full Tully-Fisher catalogue prefers a higher \(f\sigma_8\) value. Nonetheless, our constraint from the redshift cut is consistent with that without the redshift cut, albeit with slightly larger uncertainty. Similar to table~\ref{tab:TF_data} and \ref{tab:FP_data}, changing \(\sigma_u\), \(\sigma_{\mathrm{cut}}\), and grid size will modify \(f\sigma_8\) mean of posterior. However, these changes are generally smaller than the uncertainty. }    
\label{tab:combined_data} 
\end{table}

Next, we apply our method to the combined data catalogue and the results are summarised in table~\ref{tab:combined_data}. The constraints are consistent with the table~\ref {tab:FP_data} since the majority of galaxies in the combined catalogue are Fundamental Plane galaxies. Similarly, applying the redshift cut on the Tully-Fisher catalogue only slightly reduces \(f\sigma_8\). Similar to table~\ref{tab:TF_data} and \ref{tab:FP_data}, changing \(\sigma_u\), \(\sigma_{\mathrm{cut}}\), and grid size will modify \(f\sigma_8\) mean of posterior. However, these changes are generally smaller than the uncertainty. Tables~\ref{tab:TF_data}, \ref{tab:FP_data}, and \ref{tab:combined_data} all demonstrate that increasing grid size has little impact on the uncertainty on \(f\sigma_8\). For the maximum likelihood fields method, changing the grid size is similar to changing \(k_{\mathrm{max}}\) in the momentum power spectrum or \(s_{\mathrm{min}}\) in the velocity correlation function approaches because any information smaller than the grid size is removed. The linear theory shows that the peculiar velocity in Fourier space is proportional to \(\frac{1}{k}\), so it is most sensitive to large scales. Therefore, removing small-scale information does not massively reduce the constraining power of peculiar velocity. However, this is not the case for the galaxy overdensity. Using a \(30h^{-1}\mathrm{Mpc}\) grid size, \(f\sigma_8\) is unconstrained if only the galaxy overdensity data is used because the small-scale information helps to break the degeneracy between \(f\sigma_8\) and nuisance parameters. Therefore, we only show the \(f\sigma_8\) constraint with the BGS sample using the \(20h^{-1}\mathrm{Mpc}\) grid size. We also apply a galaxy overdensity cut, \(\delta_g^{\mathrm{cut}} = 10\), since our model fails at highly nonlinear scales \citep{Lai_2022}, and a higher overdensity cut will significantly increase the reduced chi-squared. This removes only 6 out of \(\sim6000\) grids, so it does not significantly affect the constraint on \(f\sigma_8\). However, it massively improves the reduced chi-squared. The constraint on \(f\sigma_8\) from the BGS density sample is consistent with the ones from the combined peculiar velocity catalogue. This allows us to combine the two data sets since there is no tension between them. 

\begin{table}[t!]
\centering                          
\renewcommand{\arraystretch}{1.4} 
\begin{tabular}{c|c|c|c|c|c|c}        
Data & $f\sigma_8$ & Reduced $\chi^2$ & dof & grid size & $\delta_g^{\mathrm{cut}}$ & $z_{\mathrm{cut}}$\\ \hline \hline
BGS  &$0.521_{-0.070}^{+0.065}$ & $1.024$ & $5929$ & $20$ & 10 & None\\
\hline
\end{tabular}
\caption{Constraints on \(f\sigma_8\) with only the BGS clustering catalogue. We only show the constraint on \(f\sigma_8\) with the \(20h^{-1}\mathrm{Mpc}\) grid size because \(f\sigma_8\) for the \(30h^{-1}\mathrm{Mpc}\) grid size is poorly constrained due to degeneracy between different parameters after removing the small scales information. }    
\label{tab:BGS_data} 
\end{table}

\begin{table}[t!]
\centering                          
\renewcommand{\arraystretch}{1.4} 
\begin{tabular}{c|c|c|c|c|c|c|c|c}        
Data & $f\sigma_8$ & Reduced $\chi^2$ & dof & grid size & $\sigma_u$ & $\sigma_{\mathrm{cut}}$ &$\delta_g^{\mathrm{cut}}$ & $z_{\mathrm{cut}}$\\ \hline \hline
all  &$0.500_{-0.030}^{+0.043}$ & $1.012$ & $10729$ & $20$ &$16$ &$4$ & 10 & None\\
\hline
all  &$0.485_{-0.054}^{+0.068}$ & $1.009$ & $3760$ & $30$ &$16$ &$4$ & 2 & None\\
\hline
\hline
$\boldsymbol{\mathrm{all}}$ & $\boldsymbol{0.483_{-0.026}^{+0.049}}$ & $\boldsymbol{1.011}$ & $\boldsymbol{10716}$ & $\boldsymbol{20}$ & $\boldsymbol{16}$ & $\boldsymbol{4}$ & $\boldsymbol{10}$ & $\boldsymbol{\mathrm{0.05}}$ $\boldsymbol{\mathrm{TF}}$ $\boldsymbol{\mathrm{cut}}$\\
\hline
all & $0.453_{-0.053}^{+0.048}$ & $0.986$ & $3755$ & 30 & $16$ &$4$ & $2$ & 0.05 TF cut \\
\hline
\end{tabular}
\caption{Constraints on \(f\sigma_8\) by combining the Tully-Fisher and Fundamental Plane peculiar velocity catalogues and the BGS galaxy catalogue. Only the statistical uncertainty is included here. The \(30h^{-1}\mathrm{Mpc}\) grid size provides a weaker constraint because removing the small-scale information weakens the constraints from galaxy overdensity. The last row in bold highlights our fiducial constraint on \(f\sigma_8\) using the maximum likelihood fields method.}    
\label{tab:BGS_PV_data} 
\end{table}

Table~\ref{tab:BGS_PV_data} presents the constraint on \(f\sigma_8\) after combining the Tully-Fisher and Fundamental Plane peculiar velocity catalogue and the BGS density catalogue. With \(30h^{-1}\mathrm{Mpc}\) grid size, adding the peculiar velocity catalogue helps to break the parameter degeneracies so that we can obtain a constraint on \(f\sigma_8\). We also apply a galaxy overdensity cut \(\delta_g^{\mathrm{cut}} = 2\) when using \(30h^{-1} \mathrm{Mpc}\) grid size. The universe is more homogeneous on larger scales, so a lower galaxy overdensity cut is required to remove the outliers. This removes 41 out of 2023 grids, which is around 2\% of the total number of grids. Therefore, this will have minimal effect on the dataset's constraining power while reducing the analysis's systematics. We decided to use \(20h^{-1} \mathrm{Mpc}\) grid size, which provides a tighter constraint on \(f\sigma_8\) due to extra information from the small-scale galaxy overdensity data. However, the constraints on \(f\sigma_8\) are consistent for the two different grid sizes, so the choice of grid size will not change the conclusion of this work. Furthermore, the constraint on \(f\sigma_8\) without cutting the high-redshift Tully-Fisher galaxies is also consistent with the one obtained with the redshift cut, since it removes less than 5\% (3872 out of 80628) of all galaxies. Therefore, the decision to exclude Tully-Fisher galaxies at higher redshifts will not change the conclusion here.

\acknowledgments

The authors would like to thank Corentin Ravoux and Christoph Saulder for providing valuable comments during the DESI internal review. YL, CH, and TD acknowledge support from the Australian Government through the Australian Research Council’s Laureate Fellowship (project FL180100168) and Discovery Project (project DP20220101395) funding schemes. YL is also supported by an Australian Government Research Training Program Scholarship and by the U.S. Department of Energy Cosmic Frontier program, grant DE-SC002599.    

This material is based upon work supported by the U.S. Department of Energy (DOE), Office of Science, Office of High-Energy Physics, under Contract No. DE–AC02–05CH11231, and by the National Energy Research Scientific Computing Center, a DOE Office of Science User Facility under the same contract. Additional support for DESI was provided by the U.S. National Science Foundation (NSF), Division of Astronomical Sciences under Contract No. AST-0950945 to the NSF’s National Optical-Infrared Astronomy Research Laboratory; the Science and Technology Facilities Council of the United Kingdom; the Gordon and Betty Moore Foundation; the Heising-Simons Foundation; the French Alternative Energies and Atomic Energy Commission (CEA); the National Council of Humanities, Science and Technology of Mexico (CONAHCYT); the Ministry of Science, Innovation and Universities of Spain (MICIU/AEI/10.13039/501100011033), and by the DESI Member Institutions: \url{https://www.desi.lbl.gov/collaborating-institutions}. Any opinions, findings, and conclusions or recommendations expressed in this material are those of the author(s) and do not necessarily reflect the views of the U. S. National Science Foundation, the U. S. Department of Energy, or any of the listed funding agencies.

The authors are honored to be permitted to conduct scientific research on I'oligam Du'ag (Kitt Peak), a mountain with particular significance to the Tohono O’odham Nation.

\section{Code and data availability}
The codes used in this paper are available at \url{https://github.com/YanxiangL/Peculiar_velocity_fitting/tree/desi}. The DESI PV DR1 catalogue will be made available upon acceptance of the papers. The data points and codes used to generate the figures in this paper are available at \url{https://doi.org/10.5281/zenodo.17602818}. 

\bibliographystyle{JHEP}
\bibliography{example.bib}



\end{document}